\documentclass{emulateapj}

\pdfoutput=1
\usepackage{epsfig}
\usepackage{amsmath}
\usepackage{natbib}
\usepackage{float} 
\usepackage{longtable}
\begin{document} 
\setcounter{equation}{0}

\newcommand{\HI}{{H~{\footnotesize I }}}
\def\kms{km s$^{-1}$} 
\def\et{\it et al.\rm}
\def\and{$\&$ } 
\def\sig{$\sigma$}


  \title{The CARMA Paired Antenna Calibration System: \\ 
         Atmospheric Phase Correction for Millimeter Wave 
         Interferometry and its Application to Mapping the 
         Ultraluminous Galaxy Arp 193}
  \author{B. Ashley Zauderer\altaffilmark{1,2}, 
          Alberto D. Bolatto\altaffilmark{1,3},
          Stuart N. Vogel\altaffilmark{1}, 
          John M. Carpenter\altaffilmark{4},
          Laura M. Per\'{e}z\altaffilmark{6,7},
          James W. Lamb\altaffilmark{5},
          David P. Woody\altaffilmark{5},
          Douglas C.-J. Bock\altaffilmark{8},
          John E. Carlstrom\altaffilmark{10},
          Thomas L. Culverhouse\altaffilmark{10}, 
          Roger Curley\altaffilmark{1},
          Erik M. Leitch\altaffilmark{5,10},
          Richard L. Plambeck\altaffilmark{9},
          Marc W. Pound\altaffilmark{1},
          Daniel P. Marrone\altaffilmark{11},
          Stephen J. Muchovej\altaffilmark{5}, 
          Lee G. Mundy\altaffilmark{1},
          Stacy H. Teng\altaffilmark{1,12},
          Peter J. Teuben\altaffilmark{1},
          Nikolaus H. Volgenau\altaffilmark{5},
          Melvyn C. H. Wright\altaffilmark{9}
          Dalton Wu\altaffilmark{1} 
  }
  \altaffiltext{1}{Department of Astronomy, University of Maryland, 
                   College Park, MD 20742}
  \altaffiltext{2}{NSF Astronomy \& Astrophysics Postdoctoral Fellow, 
                   Department of Astronomy, Harvard University, 
                   Cambridge, MA 02138}
  \altaffiltext{3}{Humboldt Fellow, Max-Planck Institute for Astronomy, 
                   Heidelberg, Germany}
  \altaffiltext{4}{California Institute of Technology, Department of 
                   Astronomy, MC 249-17, Pasadena, CA 91125} 
  \altaffiltext{5}{California Institute of Technology, Owens Valley 
                   Radio Observatory, Big Pine, CA 93513}
  \altaffiltext{6}{National Radio Astronomy Observatory, P.O. Box 0, 
                   Socorro, NM 87801}
  \altaffiltext{7}{Jansky Fellow}
  \altaffiltext{8}{CSIRO Astronomy and Space Science, P.O. Box 76, 
                   Epping NSW 1710, Australia}
  \altaffiltext{9}{Radio Astronomy Laboratory, University of California, 
                   Berkeley, 601 Campbell Hall, Berkeley, CA 94720}
  \altaffiltext{10}{Department of Astronomy and Astrophysics, University 
                    of Chicago, 5640 S. Ellis Ave., Chicago, IL 60637}
  \altaffiltext{11}{Department of Astronomy, Steward Observatory, 
                    University of Arizona, Tucson, AZ 85721}
  \altaffiltext{12}{NASA Postdoctoral Program Fellow, Goddard Space 
                    Flight Center, Greenbelt, MD 20771}

\begin{abstract}   
Phase fluctuations introduced by the atmosphere are the main limiting
factor in attaining diffraction limited performance in extended
interferometric arrays at millimeter and submillimeter wavelengths.
We report the results of C-PACS, the Combined Array for Research in 
Millimeter-Wave Astronomy Paired Antenna Calibration System. We present 
a systematic study of several hundred test observations taken during the 
$2009-2010$ winter observing season where we utilize CARMA's eight 3.5-m 
antennas to monitor an atmospheric calibrator while simultaneously 
acquiring science observations with 6.1-m and 10.4-m antennas on 
baselines ranging from a few hundred meters to $\sim$2 km. We find that 
C-PACS is systematically successful at improving coherence on long 
baselines under a variety of atmospheric conditions. We find that the 
angular separation between the atmospheric calibrator and target source 
is the most important consideration, with consistently successful phase 
correction at CARMA requiring a suitable calibrator located 
$\lesssim6^\circ$ away from the science target. We show that cloud cover 
does not affect the success of C-PACS. We demonstrate C-PACS in typical 
use by applying it to the observations of the nearby very luminous 
infrared galaxy Arp~193 in $^{12}$CO(2-1) at a linear resolution of 
$\approx70$~pc ($0.12\arcsec\times0.18\arcsec$), 3 times better than 
previously published molecular maps of this galaxy. We resolve the 
molecular disk rotation kinematics and the molecular gas distribution
and measure the gas surface densities and masses on 90~pc scales.  We
find that molecular gas constitutes $\sim30\%$ of the dynamical mass
in the inner 700~pc of this object with a surface density
$\sim10^4$ M$_\odot~$pc$^{-2}$; we compare these properties to those of
the starburst region of NGC~253.
\end{abstract} 

\keywords{Techniques: Interferometric, Instrumentation: Interferometers, 
          Galaxies: Arp 193, Galaxies: Starburst}

\section{Introduction}\label{pacsintro}
\subsection{Atmospheric Phase Fluctuations}\label{pacsproblem}
Many problems in astrophysics require attaining sub-arcsecond angular 
resolution. This resolution corresponds to the diffraction limit of a 
millimeter-wave interferometer with baselines of a kilometer or longer.  
Realizing the diffraction limit in these long baselines happens rarely 
because it requires a very stable atmosphere \citep{Carilli1999}.  
Variability of the index of refraction in the troposphere introduces 
variable time delays that, in effect, change the position of the source,
analogous to optical ``seeing'' \citep{CoulmanVernin1991,Masson1994}. At 
millimeter wavelengths, fluctuations in the refractive index are associated 
with changes in the water vapor content (wet terms) or in the air density 
and temperature (dry terms) in the troposphere over each antenna 
\citep{Lay1997b,Lay1997a}. The result of this positional jitter in 
interferometer images is that flux is scattered away from the source 
direction. Under these conditions, the peak flux density of a source is 
reduced by a coherence factor, 
\begin{equation}
{\rm C}=e^{-\sigma_{\phi}^2/2},
\end{equation}
where $\sigma_{\phi}$ is the rms of the atmospheric phase fluctuations 
\citep{TMS}.  

With improving receiver temperatures and growing interest in millimeter 
observations at the highest resolution, the importance of correcting for 
atmospheric phase fluctuations has increased. The troposphere is a limiting 
factor in the sensitivity and dynamic range unless a method of phase 
correction is used.  Phase correction is applicable to ground-based 
interferometers and space interferometry networks, for which at least one 
antenna is ground-based \citep{BeasleyConway1995,Bremer2002}. See 
\citet{Carilli1999}, \citet{Carilli1999ASP}, and references therein for a 
comprehensive review of the troposphere's effect on millimeter 
observations.  There are two primary categories of atmospheric phase 
correction:  indirect methods utilize measurements of 
water vapor content in the atmosphere via emission lines or continuum 
power, while direct methods measure phase errors via self-calibration, 
fast-switching, dual-beam, and paired antenna calibration.  Each
method has its advantages and limitations, which we briefly summarize.

\subsubsection{Indirect Determination of Phase Errors:  Water Vapor 
               Radiometry and Total Power}\label{pacswvr} 
The water vapor content in the atmosphere makes a large contribution to 
the path length variations in the troposphere. The water content can be 
measured by either observing a strong atmospheric emission line (water 
vapor radiometry; WVR) or the continuum emission of water (total power).  
WVR makes use of strong atmospheric water emission lines at 183~GHz or 
22~GHz. WVR at 183~GHz has been demonstrated to work on Mauna Kea at an 
elevation of approximately 4000~m, with the first operating radiometer 
built at the JCMT-CSO interferometer \citep{Wiedner2001}, and was chosen 
for the high elevation (5000~m) Atacama Large Millimeter Array (ALMA).  
However, the 183 GHz emission line is so strong it can saturate if the 
precipitable water vapor column exceeds 3~mm, limiting its usefulness at 
moderate or low elevation sites. The weaker 22~GHz water line is not 
saturated and has been tested at several observatories: the Owens Valley 
Radio Observatory (OVRO) millimeter array at an elevation of 1200~m 
\citep{Woody2000}, the Plateau de Bure Interferometer (PdBI) at an 
elevation of 2550~m \citep{BremerGL1996} and the Australia Telescope 
Compact Array (ATCA) at an elevation of 237~m \citep{Sault2007}.  As an
example, the OVRO 
system was demonstrated to effectively correct phases for 3~mm observations 
in good weather, although the system did not improve observations during 
typical observing conditions or at higher frequency, likely because of its 
hardware limitations \citep[e.g. room temperature amplifiers][]{Woody2000}.  
The presence of clouds is known to significantly degrade the phase 
correction performance of 22~GHz and 183~GHz WVR systems.

At frequencies away from these water lines, observations of the brightness 
temperature of the atmosphere allow a direct determination of the column 
density of water vapor \citep{bimawright}. Several observatories have 
explored the use of the
continuum emission for atmospheric calibration: the former 
Berkeley-Illinois-Maryland-Association (BIMA) millimeter array 
\citep{Zivanovic1992,Zivanovic1995}, the Institut de Radioastronomie 
Millim\'{e}trique (IRAM) 30~m telescope \citep{BremerGL1996,Bremer2002}, 
and the Submillimeter Array (SMA) \citep{Battat2004}.  Total power 
measurements frequently use the primary antenna receivers, which are more 
sensitive than separate dedicated antenna receivers often used for WVR.  
Uncertainties in systematics of the measurement and the contribution of 
atmospheric components such as liquid water droplets or ice crystals in 
clouds are hard to model or fit with precision.  

The indirect methods suffer from some limitations. First, these indirect
methods only measure the wet component, which usually dominates, but is 
not the sole contributor to the variable delay ($\Delta\tau$; see Fig.~1). 
Second, a major disadvantage is the reliance on an atmospheric model which 
has its own inherent uncertainties due to the large number of input 
variables and the precision with which atmospheric data are measured. 
Radiometers must be able to measure the water vapor to high precision to 
accurately compute the additional variable delay. To summarize, indirect 
methods of atmospheric correction work very well under some conditions, but 
are not necessarily robust to a broad range of conditions.

\begin{figure}[tp]
\centering
\includegraphics[totalheight=0.2\textheight,trim=6mm 4mm 0mm 4mm]
 {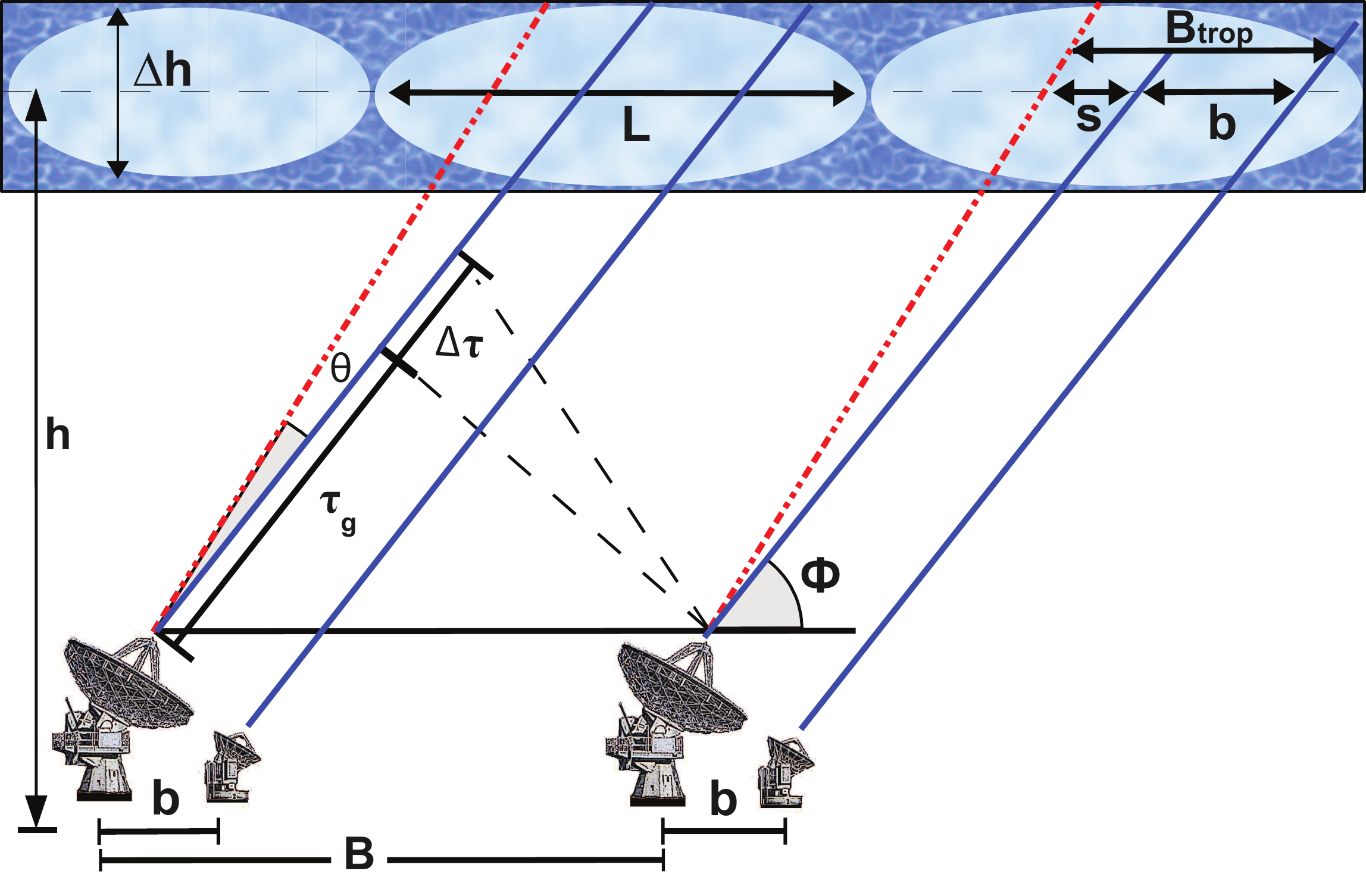}
\caption[Paired Antenna Phase Correction Geometry]%
{Atmospheric phase correction with the paired antenna method. In addition 
to the standard geometrical delay, $\tau_{g}$, water vapor fluctuations in 
the troposphere insert an additional unknown delay, $\Delta\tau$. To 
determine $\Delta\tau$, a smaller paired antenna is located near the 
primary antenna so the path through the turbulent layer will be essentially 
the same. The turbulent layer has a characteristic height, h, a thickness, 
$\Delta$h, and can be conceptualized to have an average index of refraction, 
$\it{n}$ within cells of characteristic size, L.  The paired antenna 
constantly monitors an atmospheric calibrator (solid blue) with angular 
separation, $\Theta$, from the source (dashed red). For a successful 
correction, the linear distance in the troposphere, 
$\bf{B _{\rm trop}=b + s}$, should be of order or smaller than the 
typical scale size of the turbulent cell, L (analogous to 
the size of an isoplanatic patch in adaptive optics).
}\label{diagram}
\end{figure}

\subsubsection{Direct Monitoring of Phase Errors}\label{pacsdirect}
The alternative to techniques that only measure the wet component is to
directly monitor phase errors using a point source near the target. At near 
infrared wavelengths, the adaptive optics method uses a guide star. Instead 
of a star, the radio technique uses a bright compact radio source to track 
the phase fluctuations (and associated variable delay). Instead of 
deforming a mirror in real time to apply the phase corrections, in radio
astronomy the corrections can be applied after the observations because
both amplitude and phase of the incoming wave are recorded. Regardless
of wavelength, it is important that the angular separation between the 
calibrator and source is small enough to sample the same region of the 
troposphere (see Fig.~1). Four different techniques operate on the 
principle of direct phase correction:

(1) Self-calibration.  This is a common approach in radio interferometry. 
Self-calibration requires bright, compact source structure in the
field of view, and is not broadly useful for imaging of weaker sources.
If source conditions are suitable for self-calibration, it can be
applied in conjunction with other methods
\citep{Schwab1980,CornwellWilkinson1981,CornwellWilkinson1984}.

(2) Fast Switching.  Shortening the normal source-calibrator cycle times 
can improve phase correction, but there is a trade off between time loss on 
a target source observation, and improvement made when slew times are long.  
This has motivated the development of more efficient alternatives. 
Fast-switching is implemented for ALMA ($>$84~GHz) 
\citep[see][for details]{Holdaway1992} and
for the Very Large Array in its high frequency observing modes (20$-$40 GHz) \citep{Carilli1999}.
Additionally, fast-switching at 220~GHz has been tested at Nobeyama \citep{Morita2000}. 
For fast switching, science antennas are equipped
with powerful drives which allow slewing several degrees in a few seconds.  
High sensitivity receivers are a major advantage  as this allows the
use of closer, but weaker, calibrators.  However, the atmospheric 
correction is not simultaneous with the science observation, which remains a 
major drawback. Clearly it is impractical to correct for fluctuations on 
the scale of a few seconds or shorter.

(3) Dual Beams.  In the dual-beam setup, two steerable receivers located in 
the antenna focal plane simultaneously observe sources with angular
separation ranging from 0.3 to 2.2 degrees \citep{Kawaguchi2000}. The first 
experiment was performed by \citet{Honma2003}, observing two masers at 22 
and 43~GHz. A dual-beam system has the advantage of a high sensitivity 
receiver and a stable antenna that does not need to switch between the 
target and calibrator.  One disadvantage is that the maximum angular 
separation of the beams is very limited. This limitation restricts the 
number of targets for which calibrators are available. Additionally, this 
method requires specially built and designed antennas and is not an option 
for pre-existing arrays.

(4) Paired Antenna Methods.
This technique allows simultaneous phase correction and can be implemented 
without specialized antenna designs, assuming extra antennas are available 
or can be ``borrowed'' from the primary science array. This is the method 
discussed in detail in this paper.  
We emphasize that the most important 
considerations we find for paired antenna calibration also affect fast 
switching and dual-beam calibration.

The paired antenna method for atmospheric phase correction is illustrated 
in Figure~\ref{diagram}. In addition to the standard geometrical delay, 
$\tau_{g}$, atmospheric cells (e.g. L in Fig.~\ref{diagram}) with varying 
indices of refraction, $\it{n}$, insert an additional unknown time-varying 
delay into the system, $\Delta\tau$, for antennas separated by a baseline 
distance, B.  This additional delay is related to the measured atmospheric 
phase fluctuations: 
\begin{equation}
\Delta\tau = \sigma_{\phi} / \nu_{\rm obs}~{\rm seconds}, 
\end{equation}
where $\sigma_{\phi}$ is the rms of the atmospheric phase fluctuations in 
radians and $\nu_{\rm obs}$ is the observing frequency in Hz. The paired 
antenna is placed close to the primary antenna (separation, b) so at the 
height of the turbulent layer with thickness $\Delta$h, the path through 
the atmosphere is essentially the same. The atmospheric calibrator (in the 
direction of the blue solid line, Fig.~1) is chosen with small enough 
angular separation, $\Theta$, to probe the characteristic scale size of the 
turbulence. The height of the turbulent layer can vary seasonally and 
diurnally, depending on geographic location.  
The paired antenna method works by reducing the phase fluctuations introduced
by the atmosphere from those corresponding to the physical baseline B, to
an effective baseline in the troposphere, 
\begin{equation}
{\rm B}_{\rm trop} \approx {\rm b} + {\rm s}, 
\end{equation}
where b is the physical separation between the 
science and the atmospheric monitoring antennas and s is
the additional linear separation of the antenna beams at the height of the 
turbulent layer. The linear separation, s, is minimized when the 
atmospheric calibrator is at the same azimuth as the source:
\begin{equation}
{\rm s} \approx {\rm h}/tan(\Phi-\Theta) - {\rm h}/ tan(\Phi),
\end{equation}
where h is the height of the turbulent layer, $\Phi$ is the source elevation 
and $\Theta$ is the angular separation between the source and the calibrator. 
For normal observations at moderate source elevation and a turbulent layer 
with fixed scale height, B$_{\rm trop}$ most strongly depends on the angular 
separation between the source and atmospheric calibrator, $\Theta$.
We expect the paired antenna method to reduce the atmospheric phase 
fluctuations $\sigma_{\phi}$ (corresponding to an increase in coherence, C, 
and a decrease in $\Delta\tau$) when the 
effective tropospheric baseline
B$_{\rm trop}$ is of order or smaller than the scale size, L, of the turbulent cell 
(analogous to the size of an isoplanatic patch in adaptive optics).  
The paired calibration antennas continuously monitor the atmospheric 
calibrator during science observations, so there is no loss of observing 
time and $\Delta\tau$ is well tracked.  

\begin{figure}[ht]
\centering
\includegraphics[totalheight=0.33\textheight]{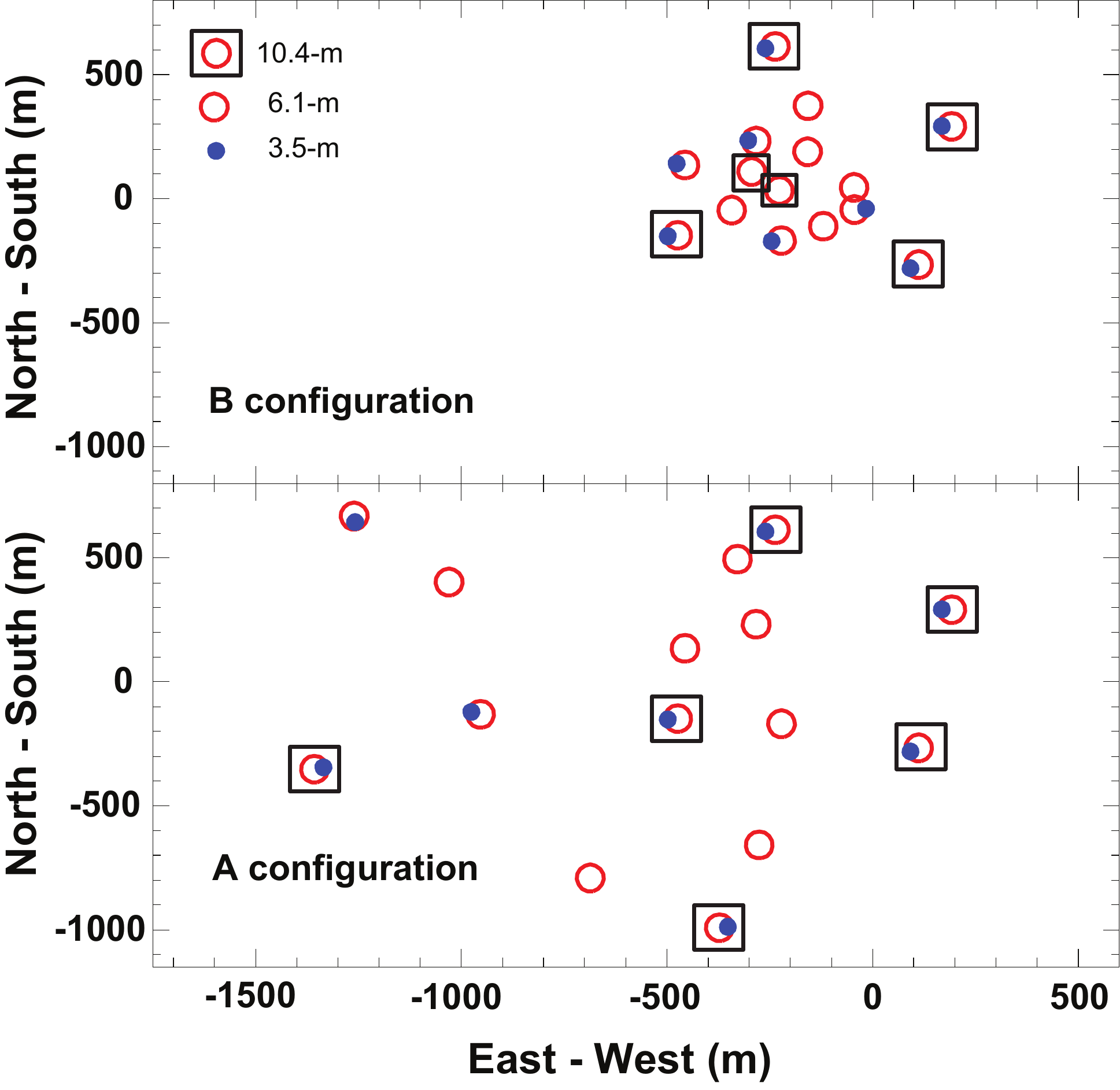}
\caption[2010 A \& B Array Antenna Configuration]%
{2009/2010 A \& B antenna configurations.  The primary science antennas 
(6.1-m \& 10.4-m antennas) are denoted by red circles with an additional
black square to indicate the 10.4-m antennas.  The 3.5-m 
paired antennas are denoted by smaller filled blue circles.  The symbols (not to scale) 
are centered on the antenna positions.  Paired antennas are positioned
$\sim$20-25 m from the science antenna. We found the paired antenna 
orientation does not affect C-PACS results. Baseline separations are 
$89 - 946$~m (B configuration) and $150-1883$~m (A configuration).}
\label{arrayConfig}
\end{figure}

Paired antenna correction was first tested at Nobeyama (NMA) by 
\citet{Asaki1996,Asaki1998}. They observed a quasar and a communications 
satellite simultaneously, using a regular science antenna for phase 
fluctuation monitoring \citep[see Figure 1 in][]{Asaki1996}. The CARMA PACS 
system (C-PACS) is unique in implementing this paired antenna phase 
correction using 3.5-m telescopes from the existing CARMA infrastructure 
with little reduction in point sensitivity. In addition, the separate 
calibration antennas can be placed close to the science antenna, and can 
observe at lower frequency, which is advantageous as most standard mm 
calibrators (e.g. quasars) are brighter at lower frequencies. The C-PACS 
experiment has eight paired baselines, for a total of 28 baselines of 
varying length and orientation. This is the largest paired antenna 
experiment to-date. \citet{Perez+10} present the first results of C-PACS, 
including the mathematical formalism and the first successful application 
to a science case. In this paper, we examine the C-PACS method 
in more detail to characterize how well the method works and under what 
conditions.

\section{Experiment Setup}\label{pacsexperiment}

We implemented C-PACS during the 2009-2010 winter observing season in 
CARMA's two longest baseline configurations, obtaining a large number of 
observations with varying angular separations between our target and 
calibrators \citep[as suggested for further work by][]{Asaki1998}.  
In the two longest baseline configurations at CARMA (A and B), we paired 
eight 3.5-m antennas\footnote[1]{The 3.5-m antennas were formerly part of 
the Sunyaev-Zeldovich Array (SZA).} with 6.1-m and 10.4-m antennas on the 
longest baselines (see Figure \ref{arrayConfig} for a graphical overview of 
the configurations).  In B configuration, four 3.5-m antennas were paired with
10.4-m antennas and four with 6.1-m antennas.  In A configuration, six 3.5-m 
antennas were paired with 10.4-m antennas, and two 3.5-m with 6.1-m antennas.
We hereafter refer to the 6.1-m and 10.4-m array of 
antennas as the ``science'' array and the paired 3.5-m antennas as the 
``calibration'' array. Infrastructure to support the calibration array was 
constructed so paired antenna pads would be as close as possible to the 
science antenna while minimizing shadowing and utilizing previous 
infrastructure constraints, such as roads and conduits for fibers. The 
distance between the paired calibration antenna and the science antenna 
ranges from 20 to 25 meters. Each array has its own local oscillator and 
correlator.  Our C-PACS tests were conducted with the science array
tuned to a sky frequency of 99.7~GHz, which we will refer to as 100~GHz.
The calibration array was tuned to a sky frequency of 30.9~GHz with
a correlator bandwidth of 8~GHz \citep{muchovej2007} centered on the
sky frequency, which we refer to as 31~GHz. 

\begin{figure*}[htp]
\centering
\includegraphics[angle=0,totalheight=0.25\textheight]{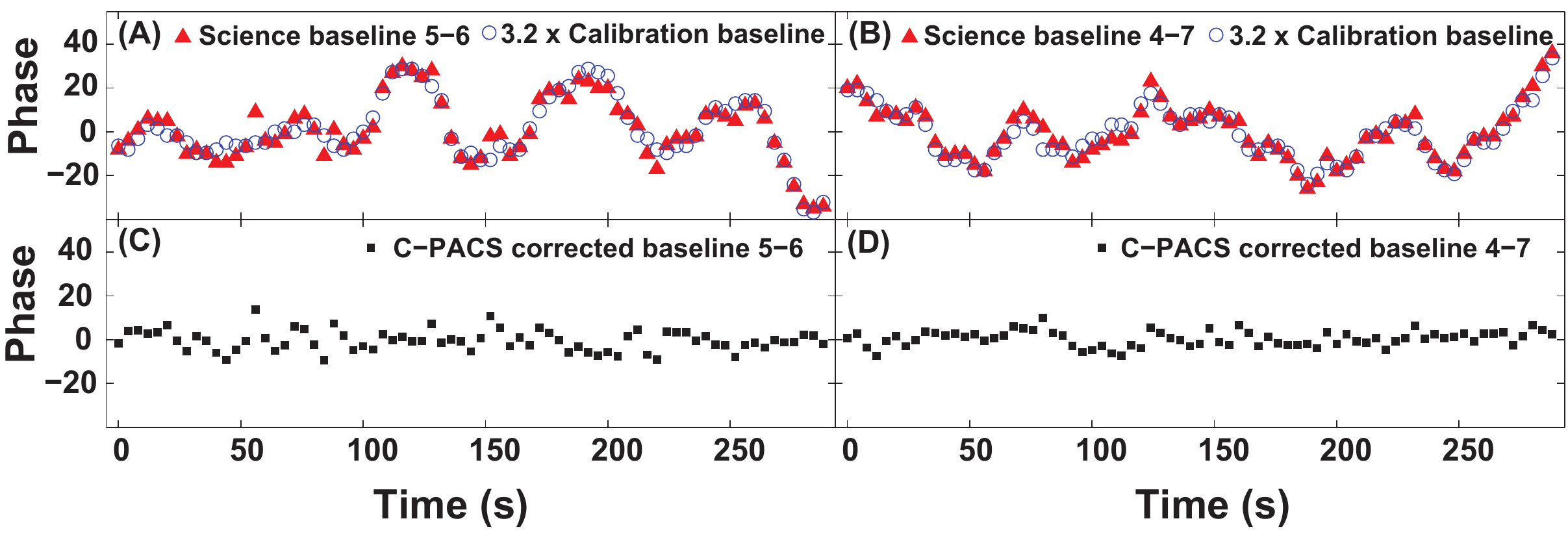}
\caption[Example C-PACS Correction]%
{
Example of C-PACS correction during A configuration obtained 2010 Jan 17 
(UT). The top panels (A, B) show the measured phases during a five minute 
observation of 3C84 for baselines 5$-$6 (1678~m) and 4$-$7 (1034~m). The 
phases for the paired antennas are scaled by the ratio of the observing 
frequencies (99.7~GHz and 30.9~GHz) because in this frequency regime the 
atmosphere is non-dispersive and the delay is the same; see equation (2).
The bottom panels (C, D) show the residual phase after C-PACS correction.  
For CARMA baseline 5-6, the rms phase decreases from 14.5$^\circ$ (A - 
red triangles) to 4.6$^\circ$ (C) after the C-PACS correction. For CARMA 
baseline 4-7, the rms phase decreases from 12.4$^\circ$ (B) to 3.5$^\circ$ 
(D). This corresponds to an improvement in coherence from 96.9\% to 99.7\%
($\Delta$C = 0.03) and from 97.7\% to 99.8\% for baselines 5-6 and 4-7, 
respectively.
}
\label{example_annotate}
\end{figure*}

To test how well C-PACS works in a variety of conditions, we designed an 
experiment to be run several times weekly. During these test observations 
(MINIPACS), the science array observes a bright source while the 
calibration array observes sources with angular separations of up to 
$\sim$12 degrees (see Table \ref{pacstable1} for properties of observed 
sources) for a duration of five minutes. An initial observation of the same 
bright source (denoted in bold; Table~\ref{pacstable1}) by both arrays 
was always included. This bright source serves as a proxy to the gain
calibrator; however, we did not return to the bright calibrator for long-time 
scale phase calibration as is standard practice every 8$-$15 minutes for
normal science observing modes.   
In total, we obtained 109 successful
MINIPACS observations in A and B configurations during the winter 
season\footnote[2]{There are seasonal variations in the mean water vapor 
content in the troposphere \citep{Bean+66}, with the lowest content 
occurring during the wintertime.} 2009-2010.  C-PACS observations were 
taken at different times each day and the final sample spans a broad range 
of observational parameters. We consider each of the 28 baselines in a 
given calibrator pair observation to be an individual ``trial". With 109 
MINIPACS observations including up to six observations of different point 
sources, our sample includes $\sim$12,500 trials. Each trial is not 
completely independent, but we separate them in this way to consider the 
effects of baseline length and orientation. For each trial, we compute 
the rms phase scatter before and after C-PACS correction, calculate the 
corresponding coherence given in equation (1), and compare the relative 
change in coherence, $\Delta$C as described for the example trials in 
Figure~\ref{example_annotate}. 

\begin{deluxetable}{llllll}  
\tabletypesize{\scriptsize}
\tablecolumns{7}
\tablewidth{0pt}
\tablecaption{Observed Sources}
\tablehead{   
  \colhead{Source} &
  \colhead{Alias} &
  \colhead{R.A.} &
  \colhead{Dec} &
  \colhead{{\it S}$_{\rm 3mm}$} &
  \colhead{{\it S}$_{\rm 1cm}$} \\
  & & \colhead{[J2000]} & \colhead{[J2000]} & \colhead{[Jy]}& \colhead{[Jy]}
}
\startdata
J0303+472 & \nodata & 03:03:35.2  & 47:16:16.3  & 0.7 & 0.8 \\
J0310+382 & \nodata & 03:10:49.9  & 38:14:53.8  & 0.5 & 1.6 \\
J0313+413 & \nodata & 03:13:02.0  & 41:20:01.2  & 0.7 & 0.8 \\
{\bf J0319+415} & 3C84    & 03:19:48.2  & 41:30:42.1  & 3.9 & 13 \\
J0336+323 & \nodata & 03:36:52.0  & 32:19:48.6  & 1.6 & 2.8 \\
J0349+461 & \nodata & 03:49:18.7  & 46:09:59.7  & 0.3 & 0.6 \\
J0414+343 & \nodata & 04:14:37.3  & 34:18:51.2  & 0.3 & 0.7 \\
{\bf J0418+380} & 3C111   & 04:18:21.3  & 38:01:35.8  & 2.0 & 5.8 \\
J0423+418 & \nodata & 04:23:56.0  & 41:50:02.7  & 0.9 & 1.7 \\
J0432+416 & 3C119   & 04:32:36.5  & 41:38:28.4  & 0.3 & 1.2 \\
J0920+446 & \nodata & 09:20:58.5  & 44:41:54.0  & 1.1 & 1.9 \\
{\bf J0927+390} & \nodata & 09:27:03.0  & 39:02:20.9  & 3.3 & 7.2 \\
J0948+406 & \nodata & 09:48:55.3  & 40:39:44.6  & 0.5 & 0.9 \\
J1150-003 & \nodata & 11:50:43.9  & -00:23:54.2  & 0.2 & 0.7 \\
J1222+042 & \nodata & 12:22:22.5  & 04:13:15.8  & 0.7 & 1.1 \\
J1224+035 & \nodata & 12:24:52.4  & 03:30:50.3  & 0.3 & 0.3 \\
{\bf J1229+020} & 3C273   & 12:29:06.7  & 02:03:08.6  & 7.1 & 25 \\
J1239+075 & \nodata & 12:39:24.6  & 07:30:17.2  & 0.6 & 0.7 \\
{\bf J1256-057} & 3C279   & 12:56:11.2  & -05:47:21.5  & 15 & 17 \\
J1613+342 & \nodata & 16:13:41.1  & 34:12:47.9  & 2.6 & 4.3 \\
J1625+415 & \nodata & 16:25:57.7  & 41:34:40.6  & \nodata & 0.4 \\
{\bf J1635+381} & \nodata & 16:35:15.5  & 38:08:04.5  & 3.4 & 3.5 \\
J1637+472 & \nodata & 16:37:45.1  & 47:17:33.8  & 0.5 & 0.6 \\
J1640+397 & \nodata & 16:40:29.6  & 39:46:46.0  & 0.5 & 1.0 \\
{\bf J1642+398} & 3C345   & 16:42:58.8  & 39:48:37.0  & 3.7 & 5.5 \\
J1653+397 & \nodata & 16:53:52.2  & 39:45:36.6  & 0.7 & 1.0 \\
J2203+174 & \nodata & 22:03:26.9  & 17:25:48.2  & 1.3 & 1.3 \\
{\bf J2253+161} & 3C454.3 & 22:53:57.7  & 16:08:53.6  & 15 & 12 \\
\enddata
\label{pacstable1}
\end{deluxetable}

\section{Data Reduction}\label{pacsdatareduction}

We performed the majority of data reduction using the Multichannel Image
Reconstruction, Image Analysis and Display (MIRIAD) software package
\citep{Sault+95}. Errant data were flagged according to standard procedures, 
and small changes in delays due to thermal effects on the fiber optics 
were corrected using the CARMA linelength monitoring system. The visibility 
data were recorded every four seconds (15$-$30~s is typical for non-PACS
observations) to track atmospheric variations, which 
allow us to determine the unknown variable delay, $\Delta\tau$.  

Amplitude and phase calibration on timescales of 5 minutes allow us to
remove instrumental phase variations by referring the phase of each
array to a point-like phase calibrator.  The data were processed in
the standard way; after flagging and bandpass calibration, a 5 minute time-scale 
phase calibration was performed independently on the science and atmospheric 
monitoring arrays. This allows us to determine and remove phase drifts on 
time scales of several minutes.  Next, we performed a short time-scale 
self-calibration on the calibration array antennas, to obtain 
antenna gains on 4 to 10~s time scales. The residual phase variations 
determined using this fast self-calibration are proportional to the 
delays introduced by a rapidly varying atmosphere.

We applied the delays determined using the calibration antennas to the
science antennas using a custom MIRIAD task, GPBUDDY, now available as
part of the standard CARMA MIRIAD software distribution.
To apply the delays, we scale the observing frequency and subtract the phases 
measured for the calibration antenna from the science antenna at each 
instant in time.  Since the data were recorded for the science and 
calibration arrays using two separate correlators, we interpolate in time 
if there is a small offset in the time stamps between the datasets (we
note that offsets were never greater than fractions of a second).  
The scaling factor is required because the calibration array was tuned to a 
lower frequency (31 GHz) than the science array (100 GHz).  We verified 
that a phase scaling factor equal to the ratio of frequencies is 
appropriate, as the atmosphere is essentially non-dispersive in the 
frequency range of our observations (e.g. see also Asaki et al. 1998) 
---  hence $\Delta\tau$ does not depend on frequency.  
Since our science array correlator's bandwidth is several GHz wide, we calculate the scaling factor for
each frequency channel separately across our band, instead of using an
average frequency value for each local oscillator setting.  We did utilize an average frequency for the calibration array,
as this data was averaged over the bandwidth to increase the signal-to-noise. 

Examining the residual ``science target'' phases after the C-PACS
calibration we found that on occasion there exist residual slow phase trends.  
We found that fitting and removing a first order polynomial 
from the phase of the ``science target'' after doing the 
C-PACS correction systematically improves the results. 
We attribute these residual phase trends to an imperfect instrumental 
phase drift correction. Indeed, our atmospheric calibration antennas and 
our science antennas are different systems working as completely independent 
interferometers, each with its own correlator.  Presumably, slow systematic
drifts between the two arrays can be removed in a real science observation
by observing a common gain calibrator every 5$-$10 minutes, and hence
we assume removing any residual trends is appropriate.

\section{Results}\label{pacsresults}

\subsection{Successful C-PACS Correction}\label{pacssuccess}
We begin by showing an example of the C-PACS correction in 
Figure~\ref{example_annotate}. A five minute observation of the quasar 
3C84 was taken during A configuration on January 17, 2010.  Both the 
science array (6.1-m and 10.4-m antennas) and the paired antenna array 
(3.5-m antennas) observed the same source ($\Theta$ = 0$^\circ$).  We 
performed the data reduction described in $\S$\ref{pacsdatareduction}. The 
resulting gains are plotted in Figure \ref{example_annotate} (phase vs. 
time) for two of the 28 paired baselines. Figure \ref{example_annotate}A 
shows the visibility phase for baseline 5$-$6 (1678~m) and 
Figure~\ref{example_annotate}B shows the visibility phase for baseline 4$-$7 
(1034~m). The calibration antenna phases are scaled by the ratio of the 
observing rest frequencies on a channel-by-channel basis (see discussion in 
\S3). The bottom panels, Figures \ref{example_annotate}C and 
\ref{example_annotate}D, show the residual phase variation after C-PACS 
correction; significant improvement is evident. For science array baseline 
5-6, the rms phase decreases from 14.5$^\circ$ to 4.6$^\circ$ after the 
C-PACS correction, corresponding to an improvement in coherence from 96.9\% 
to 99.7\% ($\Delta$C = 0.03). For science array baseline 4-7, the rms phase 
decreases from 12.4$^\circ$ to 3.5$^\circ$. The other 26 baselines show 
similar improvement. 

\begin{figure*}[htp]
\begin{center}
\begin{tabular}{cc}
\includegraphics[angle=0,totalheight=0.35\textwidth,clip=true,trim=0mm 0mm 14mm 0mm
                   ]{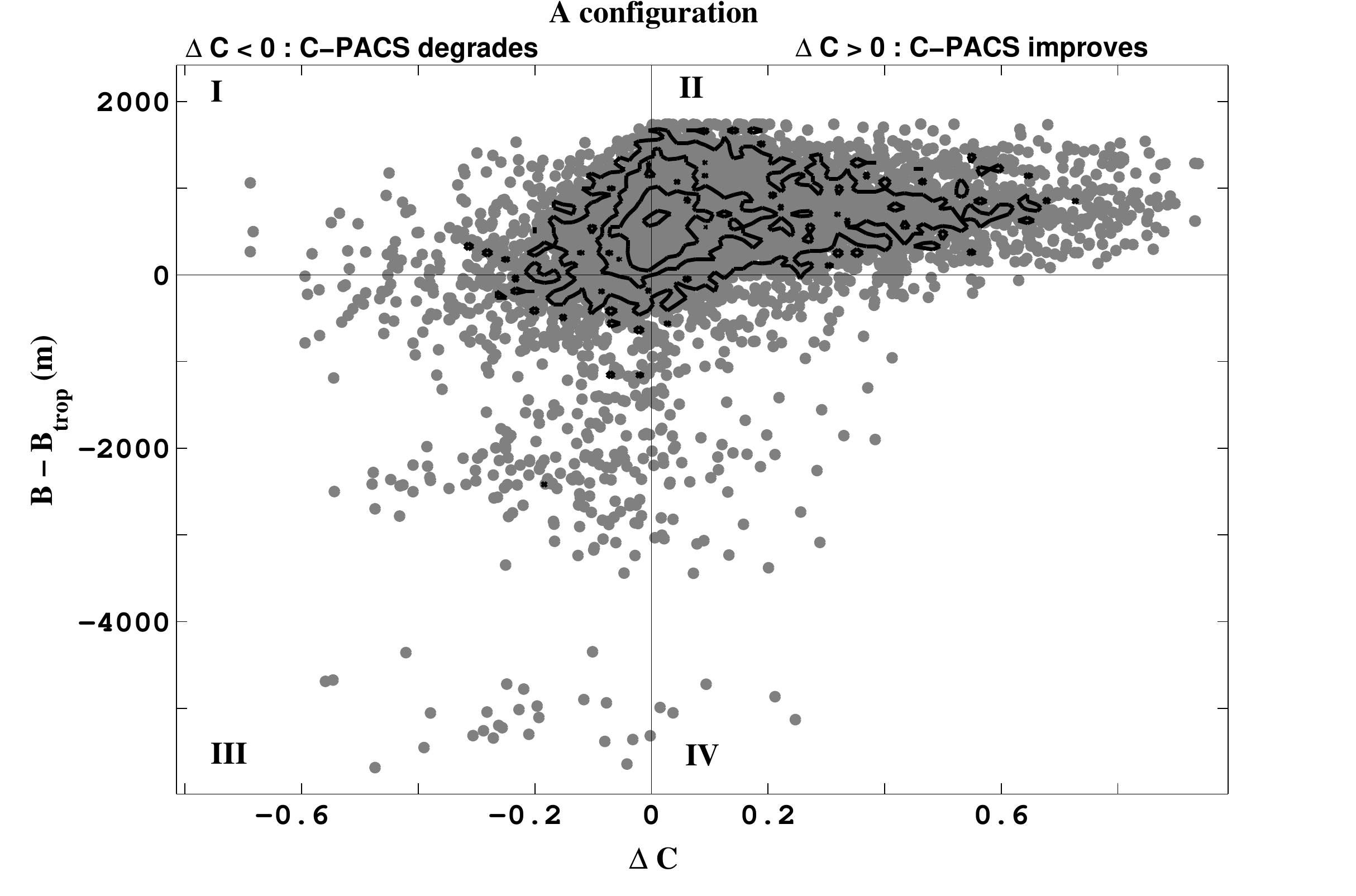} &
\includegraphics[angle=0,clip=true,totalheight=0.35\textwidth,trim=6mm 0mm 0mm 0mm
                   ]{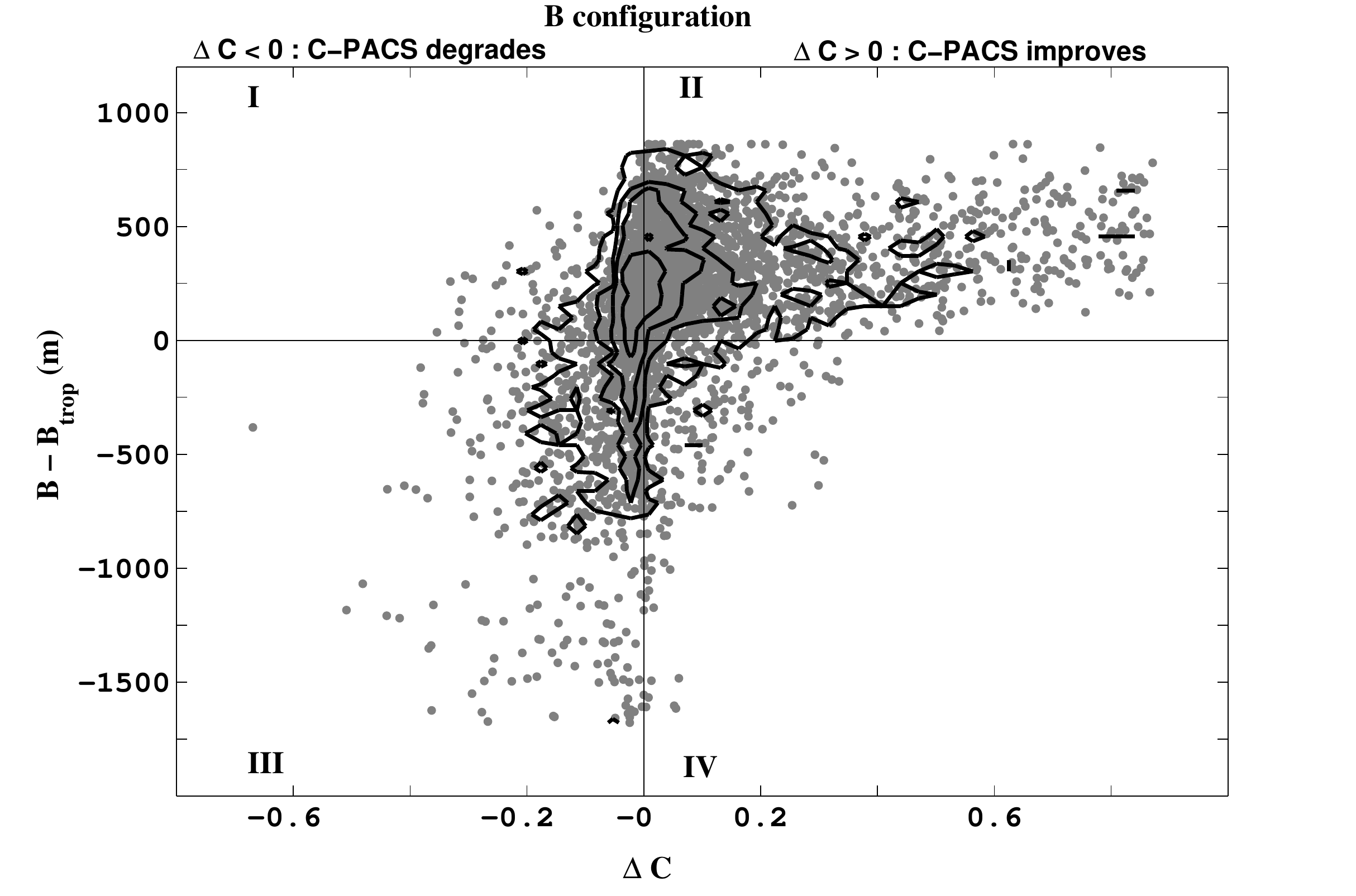} \\

\end{tabular}
\caption[Effective Baseline vs. Coherence]%
{
Change in coherence, $\Delta$C, from the C-PACS correction
as a function of B - B$_{\rm trop}$ (see Fig.~1)
for A configuration (left panel) and B configuration (right panel).
Points indicate individual baselines; density contours are overlaid at levels
3, 10, 20 and 45 points per rectangular grid cell (2500 cells per figure).
$\Delta$C is positive for a successful C-PACS correction (regions II \& 
IV).  We generally expect a successful correction for those trials
where B$_{\rm trop}$ $<$ B (see Figure 1).  There are a larger number of
shorter baselines in B configuration, explaining the larger number of
failing trials in region III (17\% for configuration B compared to 
8\% for configuration A).  In this paper we explore the trends explaining
the differences for the trials we expect to succeed (regions I and II).
}\label{beffcoh}
\end{center}
\end{figure*}

\subsubsection{Failure Modes}\label{pacsfailure}
The best way to predict if C-PACS will work during a science track is
to analyze the zero angular separation data.  For this reason, science
observations are taken on short time scales (of order a few minutes),
and bracketed with zero separation phase/atmospheric calibrator
observations. This observing setup allows systematic variations
between the arrays to be calibrated and provides a first-order check
that the C-PACS correction is working as expected.  If the zero
spacing calibration indicates that C-PACS is not working to improve
the phase calibration, then the C-PACS correction should not be used
for the interleaved observation of a science target.  After flagging
bad data, 99.6\% of our C-PACS tracks showed an 
improvement ($\Delta$C $>$ 0) for the zero angular separation data.

As angular separation between the science target and atmospheric calibrator
increase, the effective baseline in the troposphere also increases.  We
expect the C-PACS correction will be successful (improve coherence) if this effective
baseline is shorter than the actual science baseline (B$_{\rm trop}$ $<$ B).
Assuming the atmospheric calibrator is at the same azimuth as the science
target and only varies in elevation, we solve for B$_{\rm trop}$ by 
substituting equation (4) into equation (3), and taking b $\approx$25~m.
We plot the results of the C-PACS correction for A and B configurations 
in Figure~4, which includes all trials and all angular separations between
source pairs.    
The C-PACS correction is successful when $\Delta$C $>$ 0 
(i.e. Quadrants II and IV), and we expect it to be successful when 
B$-$B$_{\rm trop}$ $>$ 0 (i.e. Quadrants I and II). Hence, we are not 
extremely concerned with failures in Quadrant III, where the effective 
tropospheric baseline is larger than the actual baseline due to 
projection effects at low elevations.  Essentially, under those conditions
the atmopsheric monitoring antenna is likely sampling a very different
region of the troposphere and the correction is expected to introduce
scatter rather than reduce it.  Fig.~4 shows that C-PACS improves coherence
for the majority (70\% for A configuration; 67\% for B configuration) 
of trials for which the effective tropospheric baseline
is longer than the actual baseline.  In the remainder of this section, we
explore additional factors that lead to success (Quadrant II \& IV) or
failure (Quadrant I \& III).

\subsection{Systematic Effects}\label{pacssystematic}
In this section, we consider how angular separation, atmospheric calibrator 
flux and elevation affect the C-PACS correction for all trials shown
in Fig.~4. For successful C-PACS 
correction, the atmospheric calibrator must be close enough to the science 
target that the calibration antenna effectively samples the same 
atmospheric path, such that measured delays can be directly transferred 
to the science antenna. 

\begin{figure*}[htp]
\begin{center}
\includegraphics[angle=0,totalheight=0.5\textwidth,trim=70mm 5mm 70mm 20mm]
{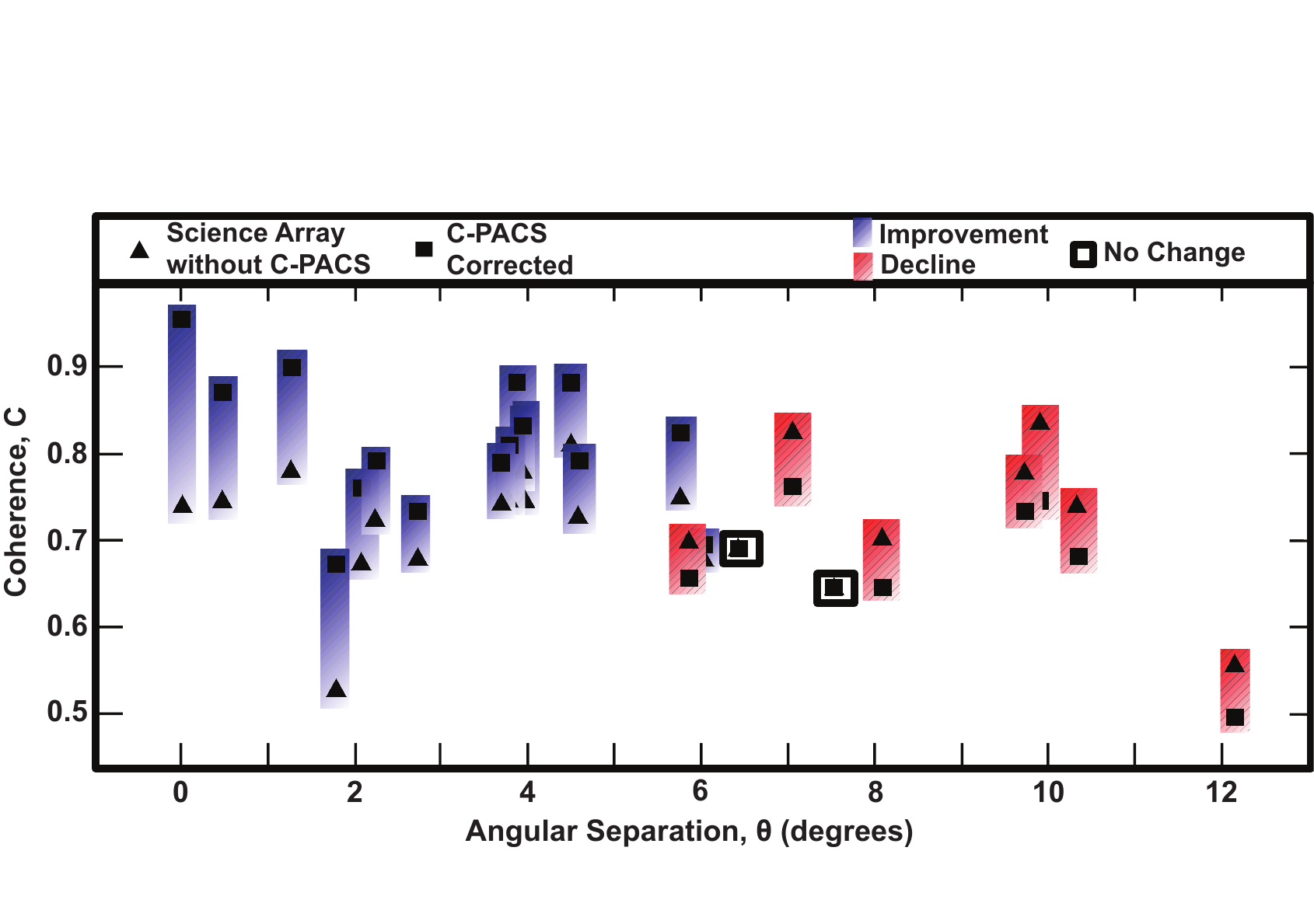}
\caption[Improvement in Coherence versus Angular Separation]%
{
Coherence as a function of angular separation.   Improvement 
in coherence after C-PACS correction (solid blue) is shown for all quasar 
pairs with $\Theta < 6^{\circ}$. For $\Theta > 6^{\circ}$, the C-PACS 
correction systematically fails and there is a decline in coherence after 
C-PACS correction (striped red).  This break at 6$^{\circ}$ suggests that 
six degrees is the typical value of the isoplanatic angle. 
}\label{angsep}
\end{center}
\end{figure*}

\begin{figure*}[htp]
\begin{center}
\includegraphics[angle=0,totalheight=0.70\textwidth]{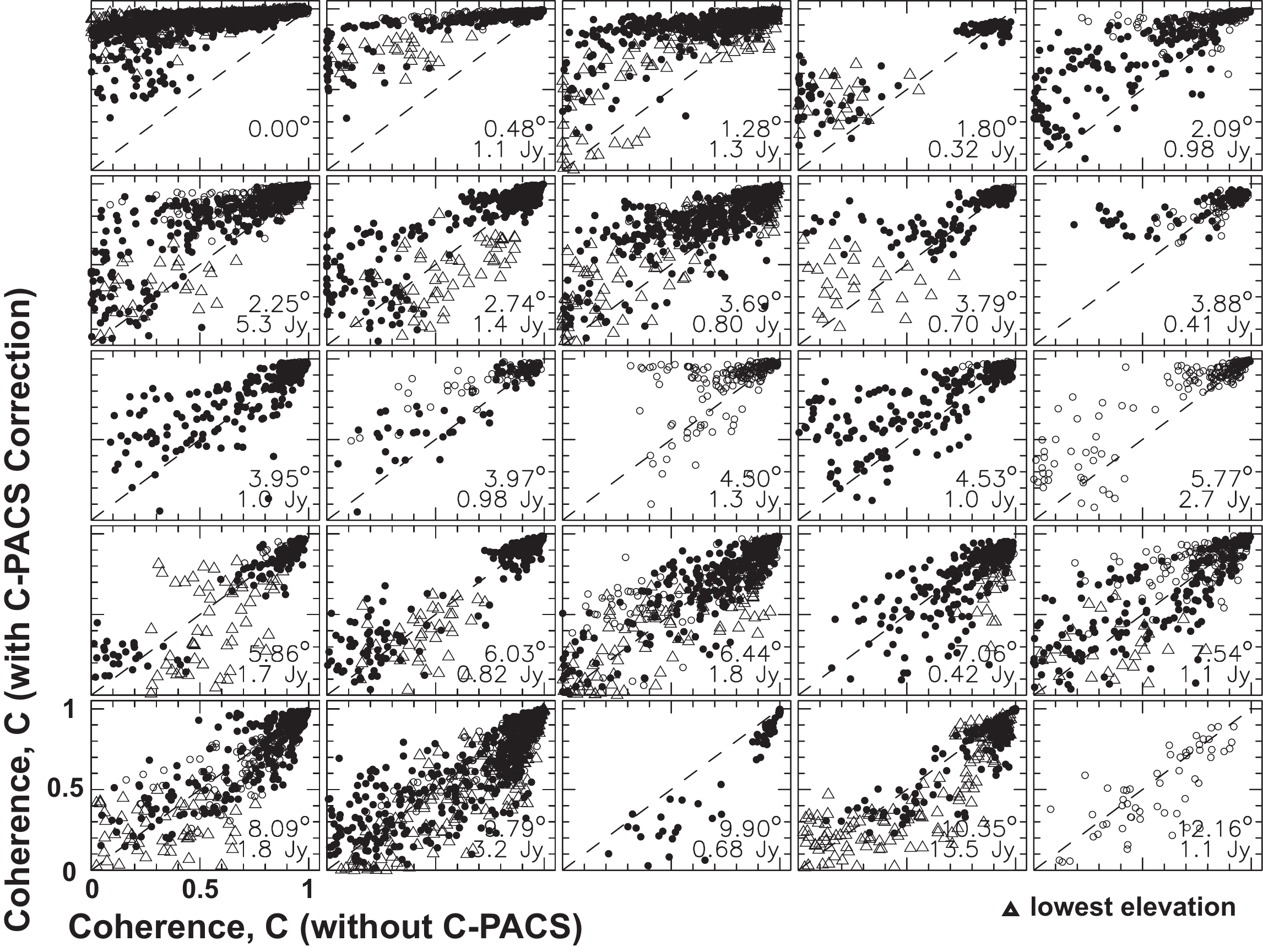}
\caption[Coh vs. coh]%
{
Coherence before (x-axis) and after (y-axis) C-PACS correction.
The diagonal line indicates the point at which the C-PACS correction would 
make no difference to the overall coherence, with points above the line 
showing improvement. Each panel represents a different calibrator-source 
pair, with the flux and angular separation noted.  The symbol shapes 
indicate elevation, decreasing from open circles ($>$65$^\circ$), to
solid black dots (35$-$65$^{\circ}$), and finally open triangles 
($\leq$ 35$^{\circ}$).

}\label{cohcoh}
\end{center}
\end{figure*}

Figure \ref{angsep} summarizes the results of the C-PACS experiment
for pairs of targets and atmospheric calibrators with different angular
separations.  We compute the average coherence before and after C-PACS
correction: the average coherence for the science data alone is denoted 
with a triangle and the average coherence after C-PACS correction with a 
square. For those angular separations where there is an improvement in 
coherence ($\Delta$C $>$ 0), we have shaded the region of improvement in 
solid blue.  For those angular separations where the coherence gets worse 
with C-PACS correction, the region is hatched and colored red. 
Figure \ref{angsep} shows that for observed sources with angular separation of less 
than six degrees between the science target and the atmospheric calibrator 
the average C-PACS correction is overwhelmingly successful, with a typical 
improvement in coherence $\Delta$ C$>$0.1, yielding an increase in peak
brightness of the observed quasar by about 15\% and a tightening of the
apparent size of the source by a few percent. For larger separations 
between science target and atmospheric calibrator the 
C-PACS correction typically fails to improve the coherence, suggesting that 
a representative isoplanatic angle for the Cedar Flat site during good 
observing conditions is 6 degrees.

\begin{figure}[ht]
\includegraphics[angle=0,totalheight=0.65\textheight,clip=true,trim=10mm 0mm 120mm 12mm
                ]{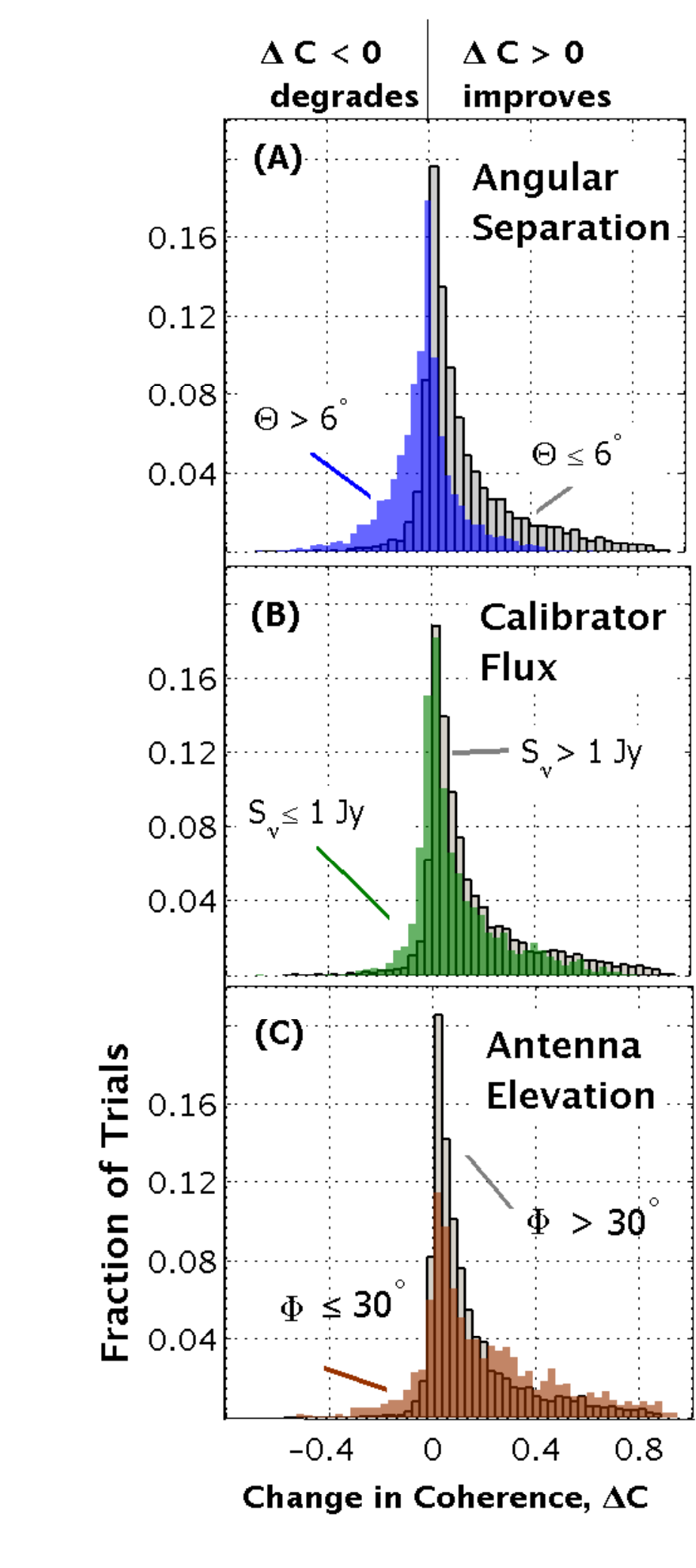}
\caption[Change in Coherence for Basic Parameters]%
{
Change in coherence ($\Delta$C = C$_{corrected}$ $-$ C$_{uncorrected}$) for 
basic calibrator parameters. Coherence is computed for every baseline in 
each track separately, as shown in Figure~\ref{example_annotate}. (A) 
Distribution as a function of angular separation, $\Theta$, between the 
calibrator and the source: 84\% of trials show improvement ($\Delta$C $>$ 
0) for $\Theta < 6^\circ$, with average improvement in 
coherence of 0.15. In contrast, only 36.5\% of trials 
show improvement for $\Theta > 6^\circ$: coherence is more 
likely to be reduced with the C-PACS correction than improved.  
For (B) and (C) we only examine trials for which $\Theta < 6^\circ$).
(B) Distribution as a function of calibrator flux.  C-PACS correction
fails more often for weaker calibrators ($\it{S} <$ 1 Jy).  
(C) Distribution as a function of calibrator elevation.  We find correction 
is successful regardless of elevation, with 82\% of trials 
showing improvement for both low and high elevations, although the 
average improvement or degradation is larger at low elevations.
}
\label{ang_flux_el}
\end{figure}

Figure \ref{angsep} summarizes the average coherence of observations, but 
in reality there is some spread in the improvement as a function of 
baseline length, source brightness, elevation, etc.  Thus, we plot for
every trial the coherence before and after the C-PACS correction for different pairs
of sources in Figure \ref{cohcoh}. The symbols 
indicate decreasing elevation of observations (open circles $>$ 65$^{\circ}$, 
filled circles 35$-$65$^{\circ}$, and open triangles $\leq$ 35$^{\circ}$).  

In Figure \ref{ang_flux_el}, we investigate the dependence of improvement 
in coherence due to C-PACS correction on angular separation, quasar flux, 
and elevation in more detail. We divide our sample into trials with angular separation 
$\Theta$ $\geq$ 6$^{\circ}$ and $\Theta$ $<$ 6$^{\circ}$ 
(Figure \ref{ang_flux_el}A). The change in coherence, $\Delta$C, is 
positive for a successful C-PACS correction.  For the $\sim$6000 trials 
with $\Theta$ $<$ 6$^{\circ}$, 84\% show improvement, with a mean 
$\Delta$C=0.15.  For the $\sim$2000 trials 
with an angular separation greater than six degrees, only 36.5\% show 
improvement.  In other words, for large angular 
separations, one is more likely to degrade observations by applying the 
C-PACS correction than to improve them.  

To evaluate the importance of the calibrator brightness 
(Figure \ref{ang_flux_el}B), we consider trials with 
angular separation, $\Theta < 6^\circ$. We bin 
our sample into two flux categories: bright (S $\geq$ 1 Jy) 
and weak (S $<$ 1~Jy). Figure~\ref{ang_flux_el}b 
shows that we systematically improve trials for the bright calibrators, 
with over 87\% showing some improvement.  The mean 
improvement in coherence is 0.18, translating to an expected 
amplitude brightening of almost 20\%. For weak calibrators,
only 65\% of the trials show improvement; however, for those which
do improve, the mean improvement is 0.15.  We note that the C-PACS 
correction is successful more often than it fails, but brighter calibrators 
produce better results more consistently.

In Figure \ref{ang_flux_el}c we show the effect of calibrator
elevation, $\Phi$ (as defined in Figure \ref{diagram}), on the
distribution of change in coherence.  We only consider trials with an angular
separation $< 6^{\circ}$. At low elevation ($\Phi \leq$ 30$^\circ$)
the same fraction, 82\%, improve as high elevation ($\Phi >$ 30$^\circ$).
However, we note that more trials at low elevation either show an improvement
or a degradation.  There are fewer trials at low elevation with little to
no change after the C-PACS correction compared to higher elevation sources. 
The impact of
elevation on the performance of the atmospheric phase correction
system is a well known phenomenon in adaptive optics, where both the
coherence length (Fried's parameter) and the isoplanatic angle depend
on the cosine of the zenith distance. Essentially, not only do the
signals travel through more atmosphere at low elevation, but the
difference in atmospheric paths tends to be greater even for nearby
calibrators, depending on the geometry. Fundamentally, as a source
moves to lower elevations in the sky it becomes increasingly difficult
for the atmospheric calibrator to sample the same portion of the
atmosphere as the science target. The effect at low elevation is
comparable to increasing the angular separation between target and
calibrator.  

\subsection{Environmental Influences}\label{pacsenvironment}

There are a large number of parameters that influence the conditions in the 
turbulent layer of the troposphere. CARMA has dedicated weather station 
equipment to measure and record air temperature, relative humidity, 
atmospheric pressure, wind speed and direction, opacity at 225 GHz, and 
atmospheric delay fluctuations.  We compute the median value of these 
weather variables for each trial and search for correlations with $\Delta$C 
after the C-PACS correction.  We single out four variables in this 
section:  atmospheric delay fluctuations, opacity, cloud cover and diurnal 
variations.  For all analysis, we only consider trials with angular 
separation less than six degrees (see previous section).  

\begin{figure}[htp]
\includegraphics[totalheight=0.7\textwidth,trim=10mm 0mm 0mm 5mm]
                {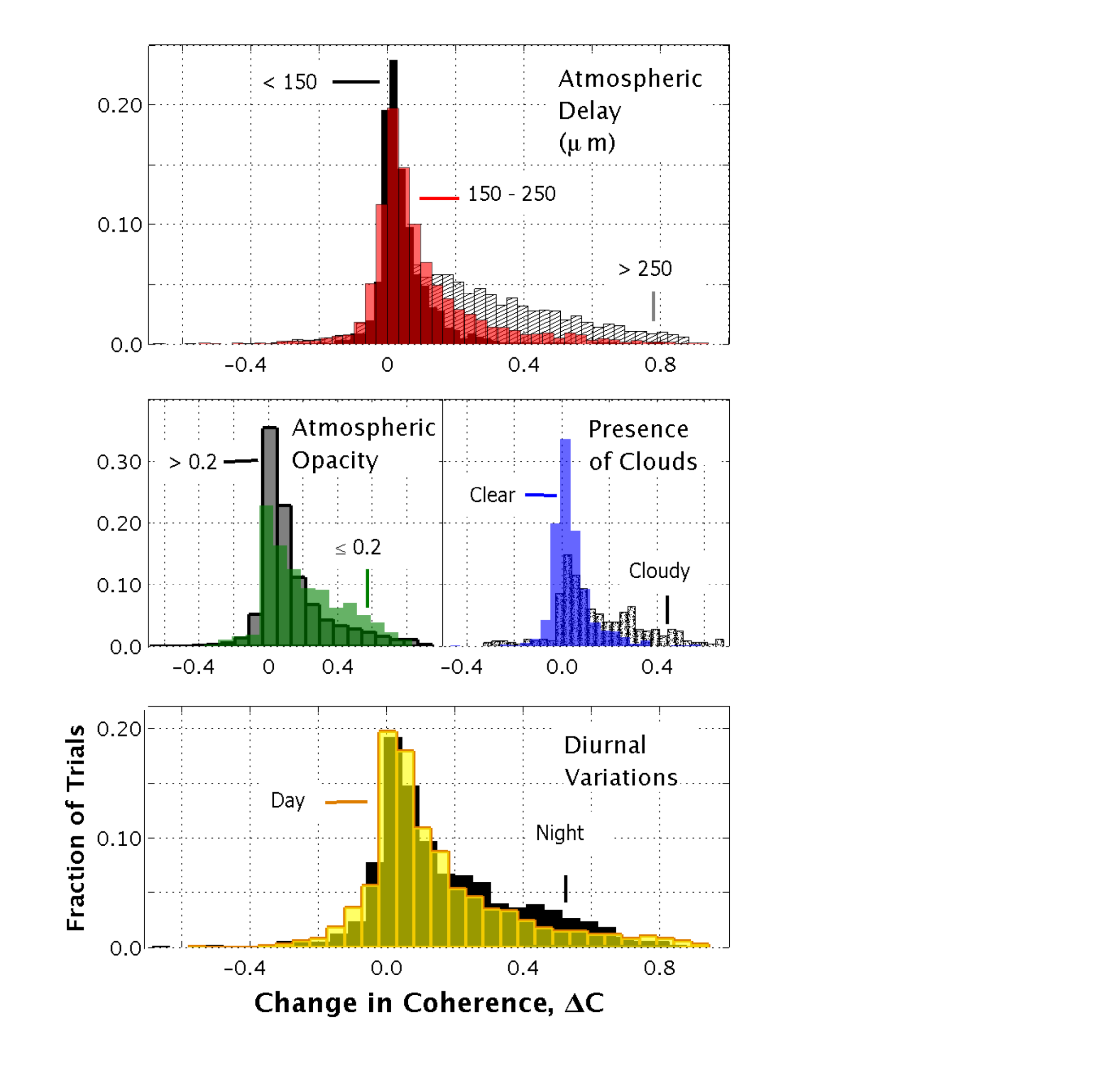}

\caption[Change in Coherence with Atmospheric Parameters]%
{
Change in coherence, $\Delta$C, after C-PACS correction for atmospheric
parameters. For the parameters examined here, we only include trials
with angular separation $< 6^\circ$. (A) Atmospheric delay. 
We find the C-PACS
correction improves coherence in weather conditions with large atmospheric
delays (c$\Delta\tau > 250 \mu$m). Coherence tends to already be high in
the best weather (c$\Delta\tau < 150 \mu$m), with only small improvement
possible with C-PACS correction. (B) Atmospheric opacity. We find the
C-PACS correction does not work as well in weather conditions with high
opacity:  C-PACS improves coherence 86\% of the time for $\tau <$ 0.2,
compared with 81\% for $\tau > 0.2$. (C) Presence of clouds. A successful
C-PACS correction is made during a period of time with cloudy conditions.
Other phase correction systems have found the presence of clouds to be a
challenge (e.g. WVR). (D) Diurnal variations. We find that coherence at
night is better to begin with, so the daytime data show a larger
improvement in coherence.  There is no major difference in the
distributions, suggesting that major characteristics of the turbulent
layer, such as height and thickness, do not significantly change diurnally.
}\label{weather_histo}
\end{figure}

The first variable we consider is atmospheric delay fluctuations.  This 
delay is measured at CARMA with a dedicated phase monitor system comprised 
of two small (18$^{\prime\prime}$) commercial antennas, forming a single 
100-m baseline.  The antenna receivers are tuned to a frequency of 
$\sim$12.5~GHz, as emitted by a geosynchronous communications satellite.  
Our ability to apply a successful C-PACS correction is not adversely 
affected when atmospheric delay fluctuations are large.  Coherence is high 
for pre-PACS data in the best weather ($\Delta\tau < 150 \mu$m), with only 
small improvement possible after applying the C-PACS correction.  We 
divide our sample into trials with large fluctuations ($>$ 250 $\mu$m), 
trials with average observing conditions (150 $-$ 250 $\mu$m), and trials 
with the most stable atmosphere ($<$ 150 $\mu$m). The distributions for 
change in coherence are shown in Figure \ref{weather_histo}a.  The C-PACS 
correction is successful in improving data in poor weather 
($>$ 250 $\mu$m): 90\% of the trials show some improvement in coherence, 
with a mean improvement of 0.28. In the very best weather, the histogram 
peaks at zero because the coherence is high (close to 100\%) without 
any correction needed: 77\% of trials 
show improvement in coherence, but the mean 
improvement is more than a factor of four smaller than in poor weather. 
In practice, phase monitor atmospheric fluctuations larger than 200 $\mu$m 
are poor conditions for observations in the high resolution A and B 
configurations.  Our results show that with a phase correction system like 
C-PACS, these weather conditions are usable.  

Next, we consider atmospheric zenith opacity ($\tau$). Zenith opacity is 
measured by a dedicated tipper, operating at 225 GHz. We have confirmed 
the accuracy of the tipper measurement with sky dips using the science 
antennas \citep{White+Zauderer08}. We bin the data into trials with 
$\tau > 0.2$ and $\tau \leq 0.2$.  Figure \ref{weather_histo}B shows that 
the C-PACS correction works well both when $\tau$ is low and high: C-PACS improves 
coherence 86\% of the time for $\tau <$ 0.2 and 81\% of the time for 
$\tau > 0.2$.  There 
is evidence that atmospheric delay and opacity are inversely related at 
other sites, such that low opacities correlate with large atmospheric 
delay fluctuations and vice versa.  If this were true, the association of 
successful C-PACS corrections with low opacity and large delay fluctuations 
would be a consequence of this correlation. We examined the measured 
opacities and delay fluctuations for each trial and find no evidence for
such inverse relation at the CARMA site.

The third environmental variable we consider is the presence of clouds.  
Numerous previous studies have concluded that other phase correction 
methods do not work reliably in the presence of clouds. Since such work 
has typically used water vapor radiometry, this is generally attributed to 
liquid and frozen water \citep{Waters1976}. We do not have equipment to 
assess cloud coverage at Cedar Flat, but we obtained weather data from the 
Western Regional Climate Center Desert Research Institute (DRI) station at 
the Bishop airport, less than 20 miles from the CARMA observatory.  DRI sky 
observations were taken hourly and include a qualitative rating of the 
cloud cover (clear, few, scattered, broken, and overcast). While we do not 
expect that there is a minute-by-minute correlation between the cloud 
coverage in Bishop and Cedar Flat, cloudy periods do tend to encompass 
large portions of the region.  Analyzing the weather data from the DRI 
Bishop airport station as a function of time shows that there are often 
several day intervals in which it is either completely clear or cloudy in 
Bishop and therefore, presumably also at the CARMA observatory.  We examine 
the distribution in $\Delta$C during one of these long extended periods of clear 
skies in Bishop, compared with tracks taken during periods of extended 
cloudy weather in Bishop (see Figure \ref{weather_histo}C).  In the case 
where there is a high probability of no clouds at the observatory site, 
over 64\% of the trials show improvement in coherence.  In the case where 
there is a high probability of it being cloudy at the observatory, 86\% of 
the trials show improvement in coherence.  We note that the mean 
improvement is $\Delta$C=0.19 for trials taken during the cloudy period and 
$\Delta$C=0.07 for trials in the clear period.  Thus, 
Figure~\ref{weather_histo}C shows that, contrary to other phase correction
techniques, C-PACS works as well in cloudy weather as in clear weather.  
This is presumably because C-PACS relies on directly measuring the atmospheric phase, 
and is not inferring it from measurements of the water vapor obtained from 
total power or spectroscopy, which may be affected by liquid
water and ice crystals.

The final environmental variable considered is time of day, motivated by
the strong diurnal pattern of wind in the north/south direction observed
in the Owens Valley \citep{Lay1997a}. To consider 
diurnal effects, we divide our sample in two by solar elevation, 
excluding sunrise and sunset when the effects of solar 
heating of the atmosphere are largest. 
The distributions of $\Delta$C are shown in 
Figure~\ref{weather_histo}D.  We find that while coherence at night is 
intrinsically better, daytime data show a slightly larger improvement in 
coherence using C-PACS (83\% of trials showing imrovement with a mean
$\Delta$C of 0.21 compared to 81\% and 0.17 for nighttime).  
We note that there is no major difference in the 
distributions, which we interpret as evidence that major characteristics of 
the turbulent layer (height, thickness, scale size of turbulent cells, outer 
scale length, wind direction and speed) do not show significant diurnal 
effects at Cedar Flat.  

In the next section, we further consider what the results of our 
observations tell us physically about the atmospheric structure.

\section{Analysis}\label{pacsanalysis}
In this section, we consider various atmospheric phase interpolation and 
weighting schemes to determine if C-PACS could be extended to nonpaired 
antennas ($\S$\ref{pacsinterpolation}). Next, we consider the effect of 
integration time on our results, specifically looking to answer how 
fast atmospheric variations occur on average ($\S$\ref{pacstimescale}).  
Finally, we discuss the predictions of turbulence theory and compute the 
root phase structure function for all baselines ($\S$\ref{pacsSF}).  
In each case, we discuss what our findings mean for the physical parameters 
of the troposphere and the implications for atmospheric correction.  

\subsection{Interpolation}\label{pacsinterpolation}
We have demonstrated thus far that the C-PACS correction is successful if 
the atmospheric calibrator is close to the ``science 
target".  Only 28 of the 105 science
array baselines have paired antennas, generally on the longest baselines.  
Maps made including the baselines involving unpaired antennas contain 
atmospheric phase errors, and therefore improvements due to C-PACS are 
significantly diluted.  This problem is especially acute for science 
targets with significant extended emission, requiring the full sensitivity 
afforded by imaging with all 105 baselines (see $\S$6).  

To mitigate this problem of phase correction ``dilution,'' we explore how 
well we can determine atmospheric phase correction by interpolating the 
phase solutions of nearby antennas.  We have written, implemented and 
tested a variety of interpolation methods in the MIRIAD program,  GPBUDDY:  
power law, Gaussian, nearest neighbor, and top hat. For each interpolation 
method, we utilize the projected ${uv}$ distances instead of physical 
distances.  The power law method weights the phase for a 
given antenna by a factor of $R^{-\gamma}$, where R is the projected separation 
between the science antenna and the calibration antenna and $\gamma$ is the 
weighting parameter.  The Gaussian method applies a weighted average at a 
given projected distance.  The top hat method equally weights all 
calibration antenna phases within a given radius and computes the average 
for the nonpaired science antenna.  The nearest neighbor algorithm simply 
uses the phase of the nearest paired calibration antenna, allowing the user 
to specify a maximum allowed distance, beyond which the science antenna 
retains its own non-corrected gain value.

We tested all the interpolation methods on one sample MINIPACS observation 
which showed excellent improvement for the paired antennas. We found that a 
successful interpolated C-PACS improvement can be made for nonpaired 
antennas in this one example and  the benefit of the correction is 
maximized using the power law interpolation method with $\gamma$=3.5 (the 
improvement was similar for indices ranging from 2-4).  We used the power 
law interpolation method and a weighting parameter of 3.5 to compute 
interpolated corrections for a subset of MINIPACS trials chosen to be 
successful for C-PACS correction of paired-paired antennas, and for which 
$\Theta <$ 6$^\circ$, $\Phi > 45^\circ$, and $\it{S}_{Jy} >$ 2.  We compute 
$\Delta$C for all baselines, and then divide the sample by baseline type:
two paired antennas (P-P), baselines with one paired antenna and one 
nonpaired (P-N), and baselines where neither antenna has a dedicated 
calibration antenna (N-N).

\begin{figure}[htp]
\centering
\includegraphics[totalheight=0.36\textwidth,trim=100mm 0mm 20mm 0mm]
                 {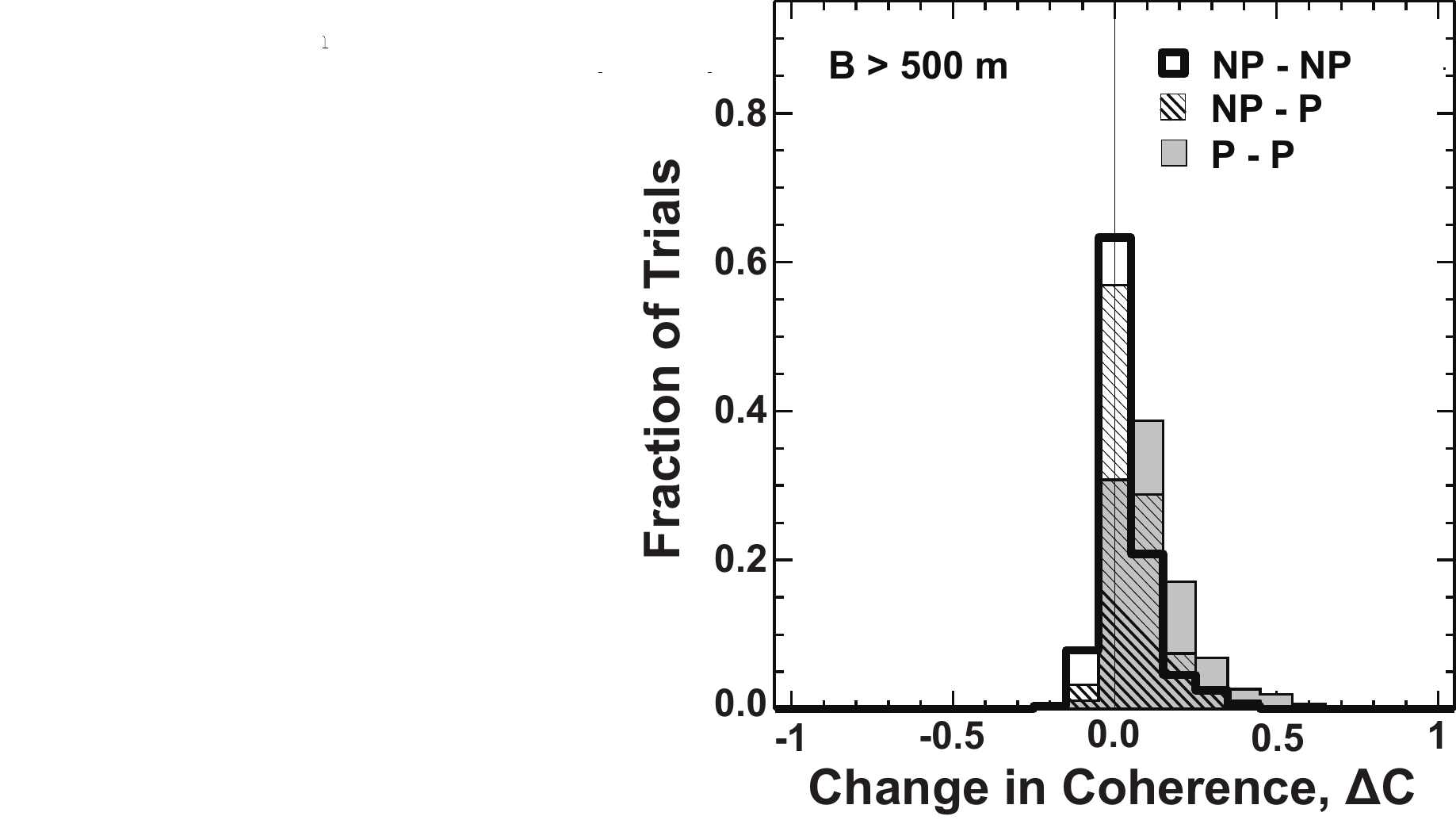}
\caption[Interpolation for Non-Paired Antennas]%
{Improvement in coherence, $\Delta$C, for interpolated baselines longer
than 500~m.  For antennas in the science array without a paired antenna, 
we compute the atmospheric correction by interpolating using a power law. 
We weight the relative contribution of gain solutions from antennas in the 
calibration array by $R^{-3.5}$.
}
\label{interpol}
\end{figure}

Figure~\ref{interpol} shows the improvement in coherence for the 
paired-paired baselines, compared to baselines with phases interpolated 
for one or both science antennas for baselines longer than 500 meters.
For the long baselines (B $>$ 500 m), 92.3\% of the P-P baselines show an 
improvement, with a median $\Delta$C of 0.10. This success rate reflects 
our choice of the best trials for this test. For long baselines with one 
nonpaired antenna, 71.4\% show an improvement in coherence (median 
$\Delta$C of 0.06). For long baselines where neither antenna had a paired 
calibration antenna, the interpolated C-PACS correction resulted in a 
success rate of 61.7\% (median $\Delta$C of 0.05).  For long baselines, we 
achieve improvement for nonpaired antennas with C-PACS, but the correction
is diluted.    For shorter baselines 
(B $<$ 500 m; not shown in Fig.~\ref{interpol}), the interpolated C-PACS correction did 
not work:  in most cases the effective tropospheric baseline is longer than the actual 
baseline (e.g. see Figure~4). The paired-paired baselines have a success rate of 74.4\% 
(median $\Delta$C of .05), nonpaired-paired baselines have a success rate 
of 53.7\% (although the median $\Delta$C of those baselines with an improvement is 0.01), 
and the nonpaired-nonpaired baselines have a success rate less than half 
(49.8$\%$, median $\Delta$C $<$ 0.01).

This experiment suggests that simple atmospheric phase correction 
interpolation dilutes the coherence improvement of nonpaired antennas to a 
significant degree, although it may be of some help for the longest 
baselines.  We think the interpolation method would work better
if the atmospheric phase screen was sampled better (i.e., more calibration
antennas).   It may also be possible to increase 
the success of interpolation by incorporating more physical 
information about the atmosphere at the time of the observations.  
Imaging the phase screen and interpolating the phases spatially and 
temporally for nonpaired antennas is an area for further investigation.

\subsection{Time Scale for Phase Variations}\label{pacstimescale}
This study used a C-PACS correction calculated with four-second
integrations.  The more rapid the atmospheric variation, the more
important it is to have fast integration times. To test how short the 
integration time needs to be in order to recover the same level of 
improvement, we did a series of tests on a sample track where there was 
excellent improvement in coherence with 4 second integrations. We 
averaged the raw data to 8, 12, 16, 20, and 30 seconds before
processing with the normal data reduction steps (flagging, bandpass, 
etc.) We then computed the coherence before and after C-PACS phase 
correction. We find that we obtain the same results with 8-12 second 
integrations, but that averaging over longer periods of time results
in a lesser improvement in coherence, and in some cases, a degradation.  
We expect these results to vary based on weather conditions and the 
strength of the calibrator as the integration time must be long enough 
to result in a strong detection of the calibrator (good signal-to-noise).
A followup investigation should be pursued as the time scale over which 
we can average and achieve improvement in coherence gives information 
about the small-scale structure of the turbulent cells in the troposphere.  
We are able to determine the thickness and outer size scale of the 
turbulent layer by computing the structure function (next section), and 
we can determine the magnitude of the small scale turbulence based on the 
integration time required to maximize coherence improvement with C-PACS 
phase correction.  

\subsection{Structure Function of the Atmosphere}\label{pacsSF}
The turbulence in the troposphere follows Kolmogorov theory 
\citep[see sections $\S$3 and $\S$4 in][]{Carilli1999}. Fluctuations 
measured by the spatial structure function, $\mathcal{D}$, correlate 
with changes in phase measured between two antennas separated by distance, 
B:
\begin{equation}
\mathcal{D}_{\Phi}(B) \equiv \langle \Phi(x + B) - \Phi(x)\rangle^2,
\end{equation}
where $\Phi$($\it{x}$) is the phase measured at one antenna, and 
$\Phi$($\it{x}$+B) is the phase measured at the other antenna in the 
baseline pair under consideration at a separation of B meters. For a single 
baseline, the ensemble average of temporal phase fluctuations are assumed 
to be equivalent to spatial fluctuations, and the measured rms phase 
variations correspond to the square root~of~$\mathcal{D}$.  
We then expect the observed behavior to follow the form
\begin{equation}
\log {\sigma_\Phi} = \log\beta + {\alpha}\log B,
\end{equation}
where $\beta$ is a scaling factor and $\sigma_{\Phi}$ is the standard 
deviation of phase scatter measured on a baseline for which a slow 
instrumental correction has been applied and atmospheric variations remain.  
As \citet{Carilli1999ASP} discuss, the scaling factor $\beta$ is the ratio 
$\frac{K}{\lambda_{mm}}$ for millimeter interferometers, where K is a 
scaling factor dependent upon the weather and $\lambda$ is the observing 
wavelength, expressed in millimeters. At excellent site locations, K has 
been found to have a typical value of $\sim$100.  It is reported that under 
good weather conditions K = 300 at the Very Large Array 
\citep[VLA][]{Sramek90}.  

\begin{figure}[htp]
\centering
\includegraphics[totalheight=0.26\textheight]{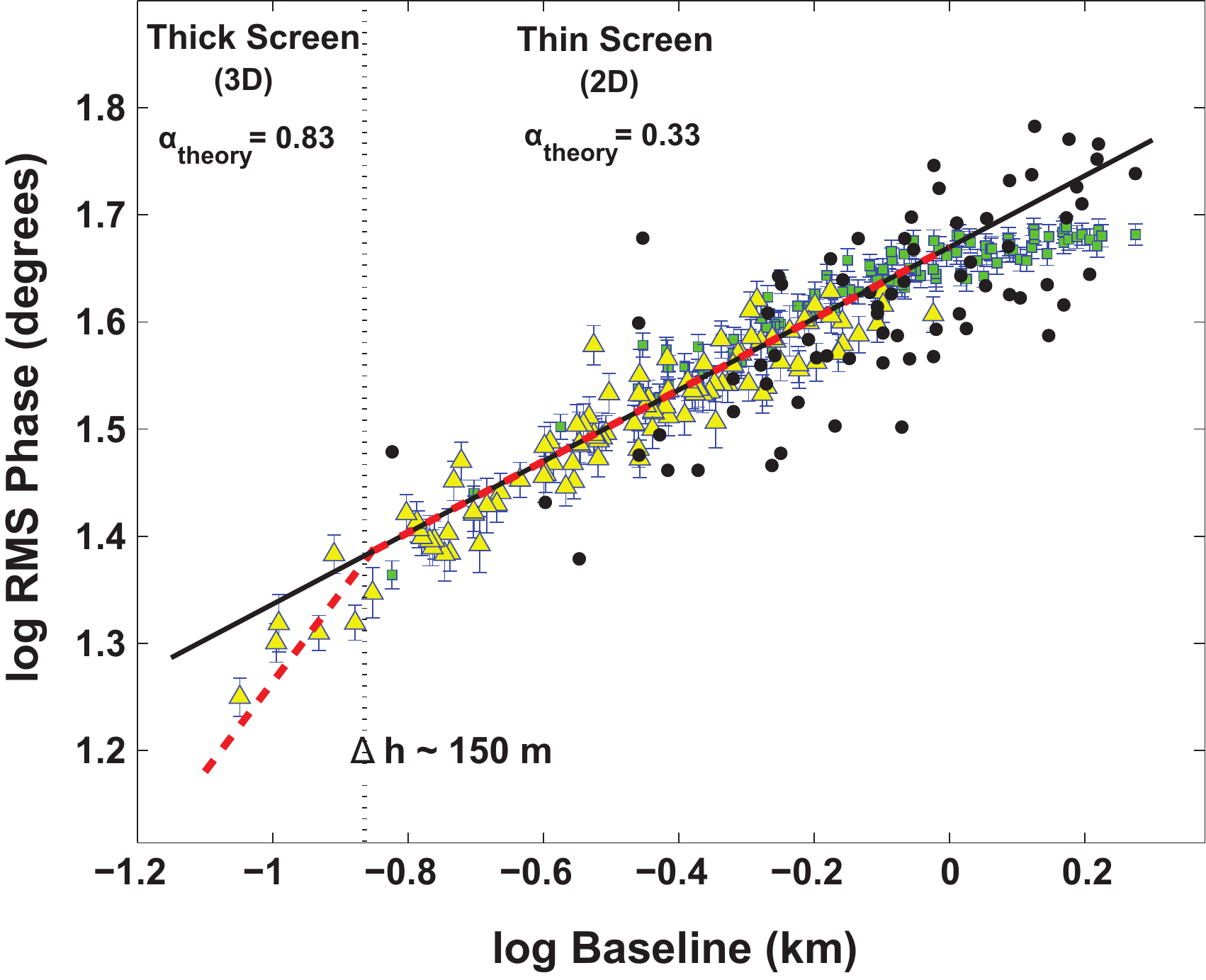}
\caption[Root Phase Structure Function]%
{
Root phase structure function for CARMA array. We bin the RMS phase scatter 
from all 15 antennas (105 baselines) by each physical baseline separation 
and plot the mean and standard deviation
for MINIPACS A Array (green squares) and B Array (yellow triangles) observations. 
The expected Kolmogorov 
power law indices of 5/6 and 1/3 for the thick and thin regimes, 
respectively,  are overlaid as slopes in this log-log plot (dashed red 
line).  The transition between these slopes suggests that the thickness of 
the turbulent layer is $\sim$150~m. According to the MINIPACS data, the 
outer scale of turbulence should be at $\sim$1 km, where the slope flattens.  
However, each MINIPACS trial was only 5$-$10 minutes in length, corresponding
to a tropospheric crossing distance of order a few kilometers.  In fact, we 
find no evidence for the outer scale to be smaller than 2~km upon 
considering a five hour observation of the phase calibrator 1310$+$323 (black points) 
during science observations of Arp 193 on February 16, 2010. The figure 
shows for the longest baselines that the theoretical slope of 1/3 is 
consistent with the data (solid black line).
}\label{structureAB}
\end{figure}

There are three scale length regimes to consider in the problem. Antenna 
baseline lengths can be longer than the thickness of the turbulent layer 
(thin screen, Kolmogorov turbulence theory predicts $\alpha$=1/3), shorter 
than the thickness of the turbulent layer (thick screen, Kolmogorov 
turbulence theory predicts $\alpha$=5/6), or the baseline length might be 
so long as to exceed the outer size scale of the turbulence. In this last 
regime, increasing the baseline length further will not increase the phase 
scatter, and $\alpha$ = 0.0. It has been found in previous studies that in 
the transition region between the thick screen and the thin screen 2-D 
approximation, the power-law index has an intermediate value.

We calculate the root phase structure function for MINIPACS experiments, 
using all 15 antennas (105 baselines) for A and B configuration.  We plot the 
mean and standard deviation of the RMS phase scatter for each baseline 
separation bin as a function of baseline separation in log-log space in 
Figure~\ref{structureAB},    
\begin{equation}
{\rm log}(\sigma_\Phi)={\rm log}(\frac{K}{\lambda}) + \alpha {\rm log}(B),
\end{equation}
to easily compute the multiplicative scaling factor and power-law index 
from a linear least-squares regression. The expected Kolmogorov power law 
indices of 5/6 and 1/3 for the thick and thin regimes are overlaid.  The 
transition between these slopes suggest that the thickness of the turbulent 
layer over Cedar Flat is approximately 150~m.  We note there are few paired 
antennas on short baselines and that this value is not well constrained.  
For the  MINIPACS data, there is a turnover to a flat slope at a baseline 
length of 1~km.  Each track was only 5-10 minutes in length, however, 
corresponding to a tropospheric crossing distance of a few kilometers 
assuming a 10~m~s$^{-1}$ wind. This suggests that the MINIPACS observations 
are too short to sample scale lengths longer than a few km, and the 
observed flattening is artificial. When we include a longer track 
(6 hours), we no longer see this clear turnover and the black points 
(Fig.~\ref{structureAB}) continue to 
follow the slope of 1/3 suggesting that the outer scale length at Cedar 
Flat is larger than 2~km. For all MINIPACS trials, we find that 
$\log\beta\approx$1.7, hence K$\approx$156 at $\lambda$=3.2 mm. This value 
of K suggests that Cedar Flat is at a location with conditions between the 
VLA (K=300) and ALMA (K=100) sites. We note that these MINIPACS trials were 
observed during the winter season with very good weather conditions which 
are not representative of the average conditions on the site throughout the 
year. We also computed the root phase structure function for the 
calibration antennas, and found the power-law index and scaling factor to 
be in good agreement with the science array for a given track suggesting 
that the calibration antennas ``see'' the same overall tropospheric 
structure as the science antennas.

\section{Science Application - Arp 193}\label{pacsscience}
In choosing a scientific case for a test of the C-PACS correction, we
considered these factors in our target selection: (1) existence of a close
($<$6$^{\circ}$) and bright ($\geq$ 1~Jy) calibrator (see \S4.2, Figures 7 
\& 8), (2) previous millimeter observations, (3) existence of comparable 
high resolution ancillary data and (4) a source with extended emission as
\citet{Perez+10} have already demonstrated dramatic improved sensitivity 
(36\% reduction in noise of image) and angular resolution (52\% decrease in 
measured size of source major axis) for the point-like science target, FU 
Orionis star PP 13S*.
  
With an interest in ultraluminous and luminous infrared galaxies
(U/LIRGs; see \S 6.1), we chose Arp 193 (also known as IC~0883, UGC~08387, VV~821,
IRAS~F13182+3424, and NVSS~J132035+340822) as the best test case for C-PACS
observations.  Unlike the closest ULIRG Arp 220, which does not have an 
appropriate calibrator within 12 degrees, Arp 193 has a nearby bright 
quasar (1310+323, 2.8$^{\circ}$ away) suitable for phase calibration and 
C-PACS atmospheric calibration according to our findings in the first part 
of this paper.  Our new maps of Arp 193 improve on the previously highest 
resolution millimeter maps by \citet[][hereafter, DS98]{DownesSolomon98} 
by a factor of $\sim3$ in angular resolution in the $^{12}$CO(2-1) line.  
Arp 193 is nearby \citep[z=0.023][]{Richter+94}, has extended emission, 
and has been studied extensively at multiple wavelengths.  Ancillary data 
is excellent for Arp 193, with these CARMA observations allowing matching 
resolution to the \HI~absorption study by Clemens and Alexander (2004) 
and optical Hubble Space Telescope (HST) NICMOS images by 
Scoville et al. (2000).

Our goal was to confirm the improvement by using the C-PACS
calibration method on an extended source. We imaged $^{12}$CO(2-1) 
in Arp 193 at sub-arcsecond scale resolution and present a brief 
analysis of the molecular gas distribution and dynamics.  We defer 
a more detailed analysis of the implications of our observations for 
a future paper.  In \S 6.1, we present a brief overview of the 
motivation to study molecular line emission in ULIRGs and summarize 
relevant scientific studies of Arp 193 and galaxies with starbursts.  
We discuss details of the observations and data reduction in \S 6.2.  
Finally, we present our results in two parts.  In the first section of 
results (\S 6.3), we discuss the success and shortcomings of the C-PACS 
phase calibration. In the second results section (\S 6.4), we analyze 
the molecular gas distribution and dynamics.

\begin{deluxetable}{lll}  
\tabletypesize{\footnotesize}
\tablecolumns{5}
\tablewidth{0pt}
\tablecaption{Science Observations of Arp 193}
\tablehead{   
  \colhead{Date} &
  \colhead{Config} &
  \colhead{Int. time (h)} 
}
\startdata
3 FEB 2007  & C (30-350 m) & 1.1 \\
14 DEC 2009 & B  (0.1-1 km)& 4.42 \\
16 FEB 2010 & A* (0.25-2 km)& 5.8 \\
\enddata
\tablecomments{*Paired Antenna Observations}\label{arptable}
\end{deluxetable}

\subsection{Background}
ULIRGs emit the majority of their energy at infrared wavelengths from
dust heated by prolific star formation (i.e. a starburst)
and/or the presence of an active galactic nucleus \citep[AGN; see][and
references therein]{2006asup.book..285L,Wilson+08}.  The only
identifying criterion for a galaxy to be classified as a ULIRG is the
measured infrared luminosity: L$_{\rm IR} >$ 10$^{11}$L$_{\odot}$ for
LIRGs and L$_{\rm IR}$ $> 10^{12}$ L$_{\odot}$ for ULIRGs.
\citet{Farrah+01} present HST observations indicating that a large fraction 
of ULIRGs (87\% in their survey) are interacting systems.  Subsequent
studies support that the majority, if not all ULIRGs, are in merging
or interacting galaxies, inferred from the disturbed morphologies,
resolved double nuclei, and tidal tails extending beyond the nuclear
region.  Due to dust obscuration of the nuclear regions
where most of the action is happening, radio observations are
critical.  High resolution imaging of CO in particular is useful for
constraining the CO-H$_2$ conversion factor, X$_{\rm{CO}}$
(DS98). \citet{Narayan+11} suggest, based on numerical simulations of
systems of merging galaxies, that not only is the conversion factor
different for merging systems than that derived for a Milky Way-like
system, but that the conversion factor can vary as a function of
radius within the disk of a merging system.

Arp 193 has a far-infrared luminosity of $4\times10^{11}$ L$_{\odot}$.  
With two clearly visible and long tidal arms, it was included in Halton 
Arp's Atlas of Peculiar Galaxies (1966). \nocite{Arp66} It is now 
understood that the narrow filaments or spikes emanating from the
nuclear region are tidal arms, evidence of a merger of two galaxies.
Arp 193 was targeted in initial studies with the Infrared Astronomical
Satellite (IRAS) and found to have higher infrared luminosity than a
control sample of noninteracting galaxies \citep{Lonsdale+84}.  The
IRAS colors ($\it{f}$$_{25}$/$\it{f}$$_{60}$ $<$ 0.2) are indicative
of cool dust \citep{Condon+Broderick91}, suggesting a starburst as the
luminosity source, rather than a central AGN.  Indeed, Arp 193 was 
categorized as a LINER\footnote[3]{Low-ionization nuclear emission-line 
region \citep[see][]{Heckman1980}.} by \citet{Veilleux+99}.  The observed
properties in LINER galaxies could arise from either low luminosity
AGNs or starbursts.  Until recently, in the case of Arp 193, the
energy source was thought to be entirely from a starburst.  However,
X-ray observations suggest the presence also of a weak AGN
\citep{Iwasawa+11,Teng2010phd}.

DS98 observed Arp 193 in the $^{12}$CO(1-0) line at 112.6 GHz
(1.6$^{\prime\prime}$ $\times$ 0.9$^{\prime\prime}$) and the
$^{12}$CO(2-1) line at 225.3 GHz (0.6$^{\prime\prime}$ $\times$
0.4$^{\prime\prime}$) between 1996 and 1998 with the IRAM
interferometer on Plateau de Bure (PdBI)
DS98 find the CO position-velocity diagram provides good
evidence for a rotating molecular ring with a minimum radius of 220 pc
and an outer disk boundary of $\sim$1300 pc based on model-fits.
Their maps suggest that the inner nuclear region hosts an extreme
starburst, similar to those in Arp 220 and Mrk 273.  These inner regions are
small ($\sim$100 pc), contain a large amount of gas mass
($\sim$10$^9$~M$_\odot$) and emit upwards of 10$^{11}$ L$_{\odot}$.

Other high resolution studies of Arp 193 include NIR and radio (\HI).  
\citet{2000AJ....119..991S} observed Arp 193 in the near-infrared with the
HST NICMOS camera, along with eight other LIRGs and 15 other ULIRGS.
Their sample includes both warm and cool galaxies (based on ${\it
f}_{25 \mu{\rm m}}/{\it f}_{60 \mu{\rm m}}$) and different types of
systems including starbursts, QSOs, Seyferts and LINERs.  The star
clusters in Arp 193 are highly luminous and hence thought to be young,
likely formed as a result of galactic interactions which are clearly
evident from the disturbed morphology of the galaxy.  In Arp 193, the
near-IR (NIR) colors are consistent with reddened starlight and a few
magnitudes of visual extinction.  \citeauthor{2000AJ....119..991S}
describe the NIR morphology of Arp 193 as a highly inclined disk.
Based on radial profile fits, they find an inner disk radius
(R$_{inner}$) of 100 pc, and an outer disk radius (R$_{outer}$) of
3800 pc for Arp 193.  They fit various models to the data, and find
the best fit is an r$^{1/4}$ law (previously recognized by Stanford
and Bushouse 1991), which suggests Arp 193 will eventually become a
spiral with a massive central bulge or possibly even a giant
elliptical galaxy.  We do not compare our data with the HST data due
to uncertainties in absolute astrometry.
\nocite{Stanford+Bush91}

Clemens and Alexander (2004) mapped the distribution of neutral
hydrogen gas in Arp 193 using the VLA and the Multi-Element Radio
Linked Interferometer Network (MERLIN).  Their high
resolution neutral hydrogen maps have a restored clean beam of
0.22$^{\prime\prime}$ $\times$ 0.20$^{\prime\prime}$.
They compare the distribution of neutral hydrogen gas with molecular
gas (CO from DS98) and near-IR HST NICMOS data.  They find that the
ISM is increasingly enriched with H$_2$ towards the center of Arp 193.
Comparing the velocity distribution of the \HI~with molecular gas,
Clemens and Alexander note variations may arise from both spatial
distribution and dynamical differences.  CARMA gives us the ability to
improve upon the molecular gas maps, achieving an angular resolution
in CARMA's A configuration that matches the HST NICMOS observations
and exceeds the \HI~MERLIN observations (which are absorption line
measurements and hence only probe near-side \HI), enabling detailed study of
the nuclear region of Arp 193 with $\sim$70 pc resolution.

\begin{figure}[htp]
\centering
\includegraphics[totalheight=0.16\textheight,trim=0mm 0mm 0mm 0mm]
                {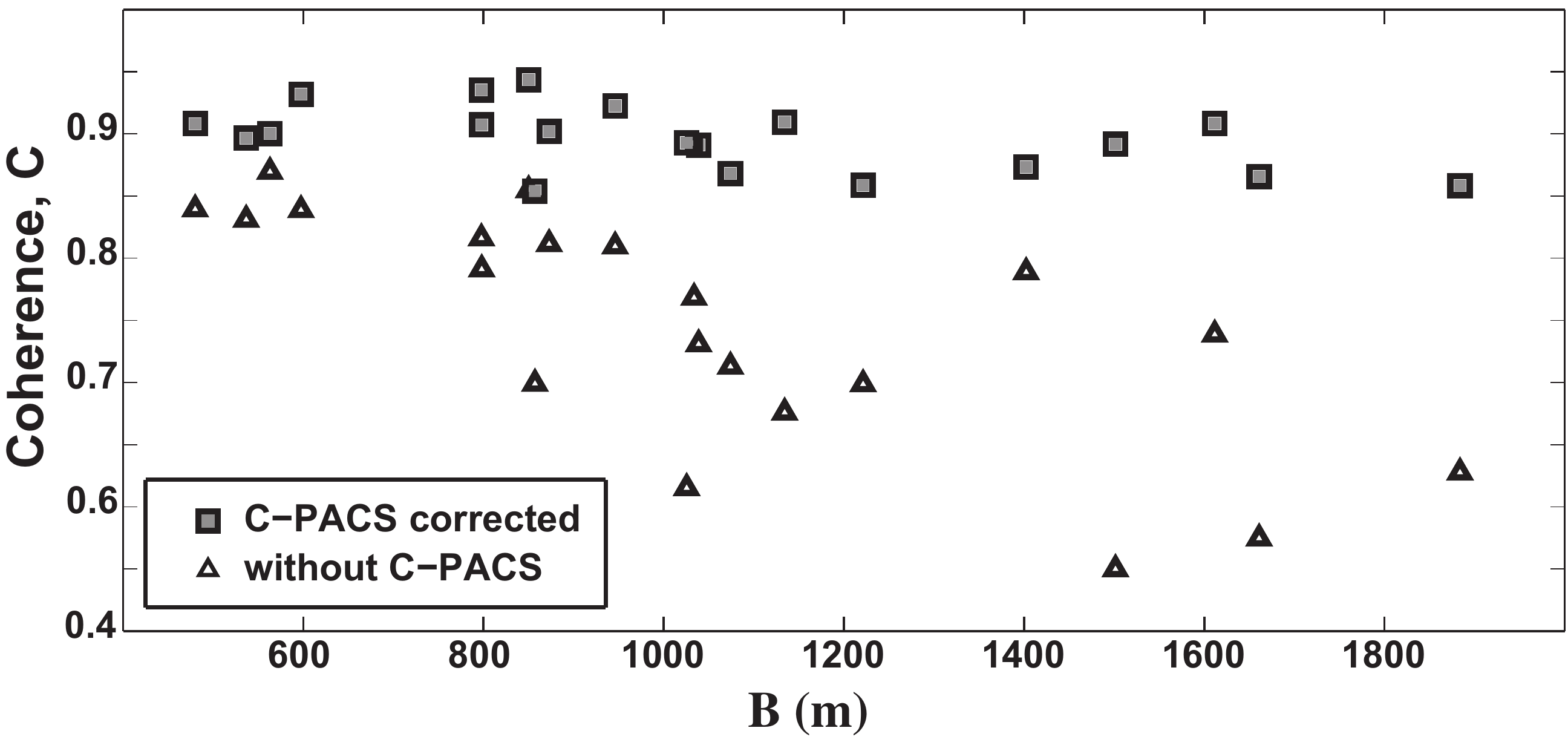}
\caption[Improvement in Coherence for Phase Calibrator 1310+323]%
{
Improvement in coherence for the phase calibrator, 1310+323, during a 
1~mm observation of source Arp 193. The mean coherence without C-PACS 
applied is 74\% and improves to 90\% with C-PACS. The improvement grows 
with increasing baseline separation, showing the importance of atmospheric 
phase correction to recover information on the longest baselines.
}
\label{1310_coh}
\end{figure}

\subsection{Observations and Data Reduction}\label{arp193data}
We observed the molecular transition $^{12}$CO(2-1) in the nuclear
region of Arp 193 in CARMA's A, B and C configurations. We summarize
the observing parameters in Table \ref{arptable}.  For all
observations, we used either 3C273 or 0854$+$201 as our bandpass and
flux calibrator, bootstrapping the flux from regular planet
measurements (the absolute flux calibration precision is $\sim$20\%). 
For the C configuration observations, we used 3C273 as the phase 
calibrator, and 1415$+$133 as a test source. We used a 14 minute cycle 
time, spending 10 minutes integrating on source, and 2 minutes on each 
of the phase and test calibrators.  For our later B and A configuration
observations, we used 1310$+$323 as the phase calibrator (2.8$^{\circ}$ 
from Arp 193) and 3C286 as a test calibrator (4.8$^{\circ}$ from 
1310$+$323).  We shortened our cycle time to 5 minutes, spending 
3 minutes on source, and one minute on each of the phase and test 
calibrators.  For A configuration C-PACS observations, 1310$+$323 was 
also the C-PACS atmospheric calibrator.

All observations were performed at 1~mm, with the observing frequency
set to 225.0483 GHz to center the $^{12}$CO(2-1) line in the lower
sideband, as Arp 193 has a redshift of $z$=0.023.  At the time of our
observations, the CARMA correlator had six windows which could be
configured to widths of 512, 64, 32 or 8 MHz.  We used the wideband
512 MHz correlator setup to accommodate the full width of the line
velocity in one window.  This resulted in a velocity resolution of
41.6~km~s$^{-1}$ per channel and an overall coverage of $-290$ to $+290$
km~s$^{-1}$ in the lower sideband. In A configuration, the atmospheric
calibrator, 1310$+$323, was observed by the calibration array at
31~GHz, as described in $\S$\ref{pacsexperiment}.  Data reduction was
performed using the MIRIAD software package to apply standard
interferometric calibrations.  C-PACS phase correction was then
applied to the A configuration data using the method described in
$\S$\ref{pacsexperiment}.  We used a power law scaling with an
exponent of 3.5 to interpolate the phase correction for nonpaired
antennas (see $\S$5.1).  All figures with maps showing relative offset
in arcseconds are with respect to the position ($\alpha$ (J2000) =
13:20:35.3 and $\delta$ (J2000) = 34:08:22.0).

\begin{figure*}[htp]
\centering
\includegraphics[angle=0,totalheight=0.35\textheight]
                   {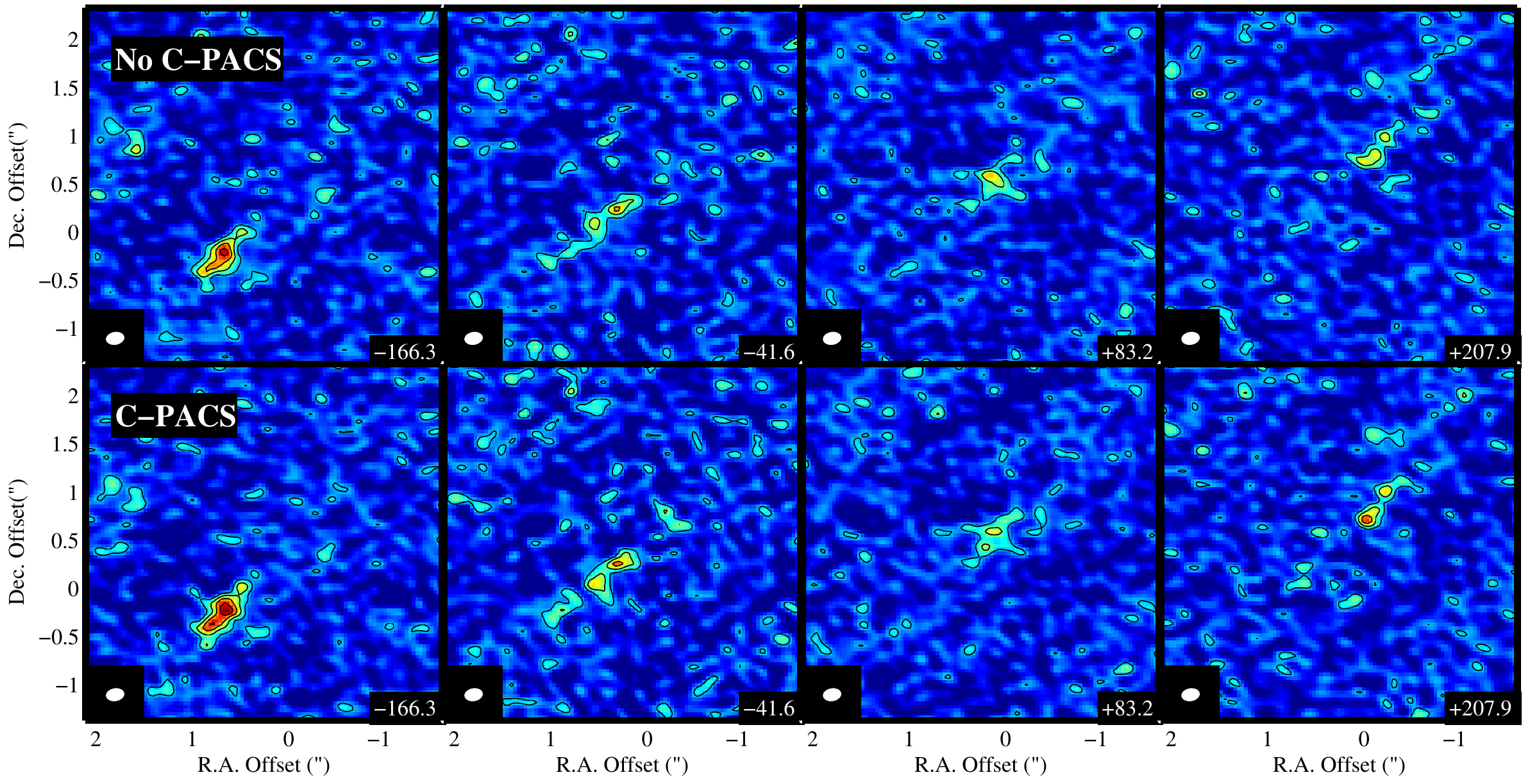}
\caption[A Array Map of Arp 193]%
{ Improvement in coherence for Arp 193 with
application of C-PACS.  $^{12}$CO(2-1) emission in 125 km s$^{-1}$
width channels is shown for data reduced without C-PACS (top panels)
and with C-PACS phase correction (bottom panels).  Contours are
plotted at 1.5, 3, 4.5 and 6 $\sigma$, where $\sigma$ = 5.4 mJy
beam$^{-1}$.  The center velocity of each channel is labeled (bottom
right, km s$^{-1}$).  Beam (lower left) is 0.18$^{\prime\prime}$
$\times$ 0.12$^{\prime\prime}$ or $\sim$84 $\times$ 56 pc. Positional
offsets are relative to the map center $\alpha$ (J2000) =
13:20:35.3 and $\delta$ (J2000) = 34:08:22.0).}
\label{arp193_A}
\end{figure*}

\subsection{Results:  Application of C-PACS}\label{arp193results}
Analysis of the phase and test calibrator data gave us confidence that
the C-PACS phase correction will result in an improved map of Arp 193.
For our A configuration observations, we applied the C-PACS correction
from observations of 1310$+$323 at 31~GHz by the atmospheric
calibration array to a test point source observed by the science
array.  We included a test source, 3C286, with an angular separation of
4.8$^\circ$ from the 1310+323.  
Applying the C-PACS phase correction from observations of 1310$+$323
by the calibrator array to the science array observations of
1310$+$323 at five minute intervals throughout the track resulted in
significant improvement.  Figure \ref{1310_coh} shows the change in
coherence for the phase calibrator, 1310$+$323.  The mean coherence
without C-PACS applied is 74\% and improves to 90\% with C-PACS.
Improvement increases with increasing baseline separation and is
striking for baselines longer than 1~km.  We did not use this information
to vary the gains in our data reduction of our science source, Arp 193, but
note that this correction would further increase the overall flux.
 
Arp 193 is situated $2.8^\circ$ away from the atmospheric calibrator, 
midway to our test source 3C286
(4.8$^\circ$ away). The mean coherence improvement in the latter is 
very small (from 46\% to 50\%). 
We expect the improvement in our
science observations of Arp~193 to be significantly better for the longer baselines 
and somewhere in between these results for zero
and 4.8$^\circ$ separation.  We note that the improvement
for the test source at 4.8$^\circ$ is smaller than expected from our MINIPACS
results (e.g. Fig.~5; $\sim$70\% to $\sim$80\%) because these science observations were 
executed at 225 GHz and the larger scale factor (7.4 compared to $\sim$3.2) 
between this 1~mm observing frequency 
and the calibration array at 31~GHz magnifies any imperfect phase measurements.

\subsection{Results: Arp 193}\label{scienceresults}
In this section, we present our $^{12}$CO(2-1) maps of Arp 193.  We
clearly resolve clumps of emission spatially and dynamically.  We
present measurements of these clumps (luminosity, mass, column density
and surface density; see Table~3) and compare the implied molecular gas mass with
the dynamical mass derived from the rotation curve we fit to our data.
At the redshift of Arp 193 (z=0.023), 1$^{\prime\prime}$
corresponds to 470~pc ($D_A\approx96$~Mpc). Its luminosity distance
is $D_L\approx98.9$~Mpc.

\begin{figure}[htp]
\centering
\includegraphics[totalheight=0.25\textheight]
                 {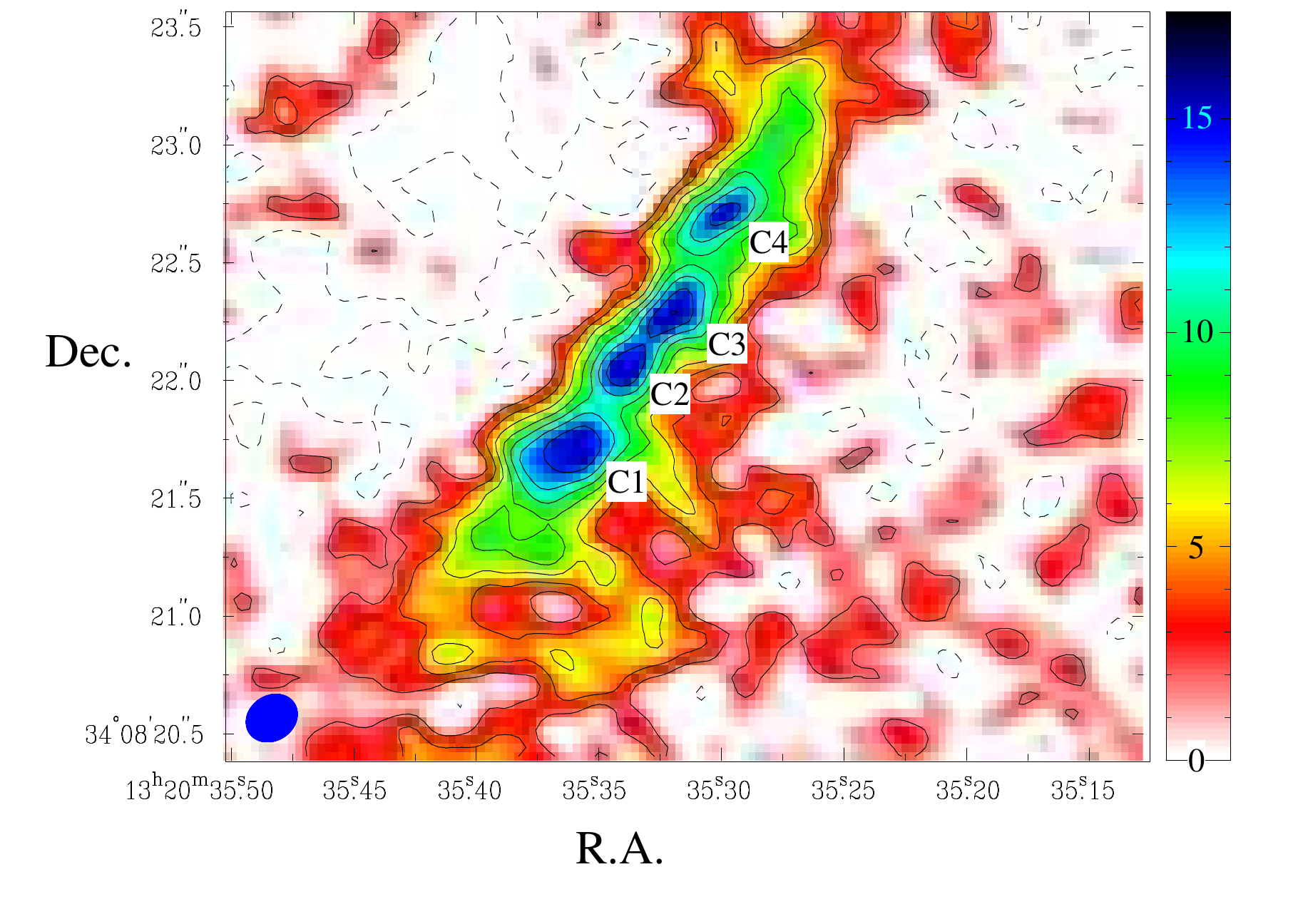}
\caption[Arp 193 Integrated Intensity Map]%
{ Integrated intensity map of $^{12}$CO(2-1) in Arp 193, using
observations from combined A, B and C configuration.  Contours are at
levels of -2, 2, 4, 6, 8, 10, 12, 14 and 16 Jy km s$^{-1}$ beam$^{-1}$
and the colorbar scale has the same units.  The rms noise in the map
is 1.74 Jy km s$^{-1}$ beam$^{-1}$.  The beam size is
$0.23^{\prime\prime} \times 0.16^{\prime\prime}$ ($107\,{\rm
pc}\times75\,{\rm pc}$), shown in the lower left .
}
\label{arp193_mom0_clumps}
\end{figure}

\subsubsection{CO Maps}
First, we present channel maps of $^{12}$CO(2-1) emission for Arp 193
using only data from the most extended configuration of CARMA (see
Fig. 2), yielding the highest resolution map.  Figure
\ref{arp193_A} illustrates the improvement in coherence achieved with 
application of C-PACS. $^{12}$CO(2-1) emission is averaged over three 
channels ($\Delta v$=125~km~s$^{-1}$)  and images are presented for 
data reduced without C-PACS (top panels) and with C-PACS phase 
correction (bottom panels). Contours are plotted at 1.5, 3, 4.5 and 
6 $\sigma$, where $\sigma$ = 5.4 mJy~bm$^{-1}$.  The center velocity 
for each map is shown in the bottom right (km s$^{-1}$).  The angular 
resolution of these maps is $0.18\arcsec\times0.12\arcsec$ equivalent 
to $\sim84{\rm pc}\times56{\rm pc}$, an improvement by a factor of 
$\sim3$ over the previous highest resolution CO map of Arp 193
\citep{DownesSolomon98}.

In Figure \ref{arp193_mom0_clumps}, we present the $^{12}$CO(2-1) 
integrated intensity map of Arp 193 using a combination of A, B and C 
configuration observations.  The data were inverted using robust 
weighting, and cleaned with a mask derived from C configuration 
observations. Because we include information from more compact 
configurations in an effort to better recover extended flux, the 
resolution of this image is slightly lower (0.23$^{\prime\prime}$ 
$\times$ 0.16$^{\prime\prime}$ or $\sim$90 pc). The total detected
flux we report out to 3-$\sigma$ significance (430.1$\pm$9.1 Jy km s$^{-1}$; Table 3)
is consistent with the total of 450 Jy km s$^{-1}$ reported by Downes \& Solomon (1998).

\begin{figure*}[htp]
\centering
\includegraphics[totalheight=0.20\textheight]{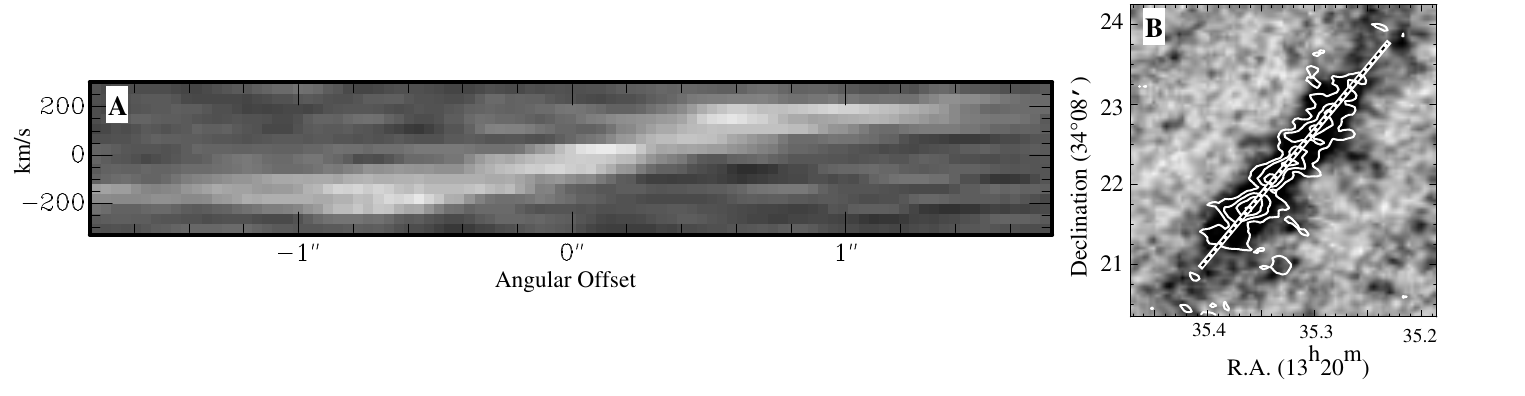}
\caption[Arp 193 Position Velocity Diagram]%
{
Arp 193 position-velocity map (A) along slice (PA=53$^{\circ}$) 
indicated (panel B).  The (0,0) position corresponds to the center
of the map ($\alpha$=13:20:35.5, $\delta$=34:08:22.0).  An angular offset 
of zero roughly corresponds to Clump C3 (Fig.~13), however the dynamical
center (see Fig.~15) is slightly closer to Clump C2.     
}
\label{arp193_posvel}
\end{figure*}

\subsubsection{Dynamics}
We summarize the dynamical information from our maps and compare with 
$^{12}$CO(2-1) images by DS98 and with \HI~maps by Clemens and Alexander 
(2004). Arp 193 is thought to be a rotating ring, inclined by 50$^{\circ}$ 
(DS98). We examined velocities along the position angle slice indicated in 
Fig.~14, using our combined A+B+C configuration map and find results consistent 
with DS98.  The position angle of the disk or ring is about
140$^{\circ}$ (E of N) and the center of rotation is coincident with Clump 
C3 (see Figure \ref{arp193_mom0_clumps}).  The coordinates of the dynamical 
center are approximately $\alpha$ (J2000) =13:20:35.318 and $\delta$ 
(J2000) =34:08:22.35.

\begin{figure}[htp]
\centering
\includegraphics[totalheight=0.25\textheight]
                {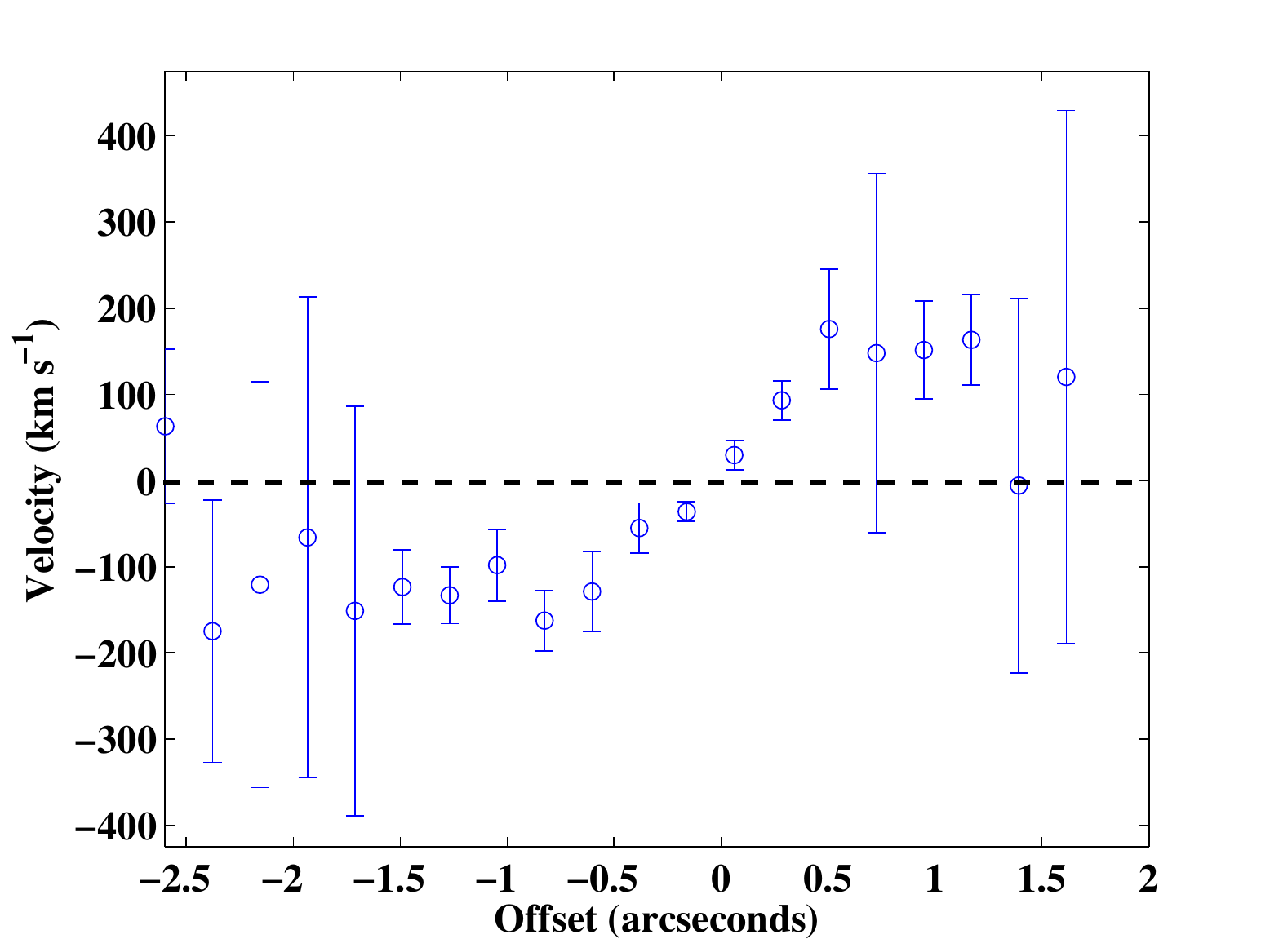}
\caption[Arp 193 Rotation Curve]%
{ 
$^{12}$CO(2-1) rotation curve for the position angle indicated in 
Fig.~\ref{arp193_posvel}B.  The slice we used to obtain this rotation curve
is thicker, averaging over two beam widths ($0.4\arcsec$) along the minor 
axis to incorporate the full thickness of the emission and improve 
signal-to-noise. The velocity and error bars at each point along this slice were determined 
by running a Monte Carlo simulation (1000 trials) whereby we randomly added white noise 
and then fit a gaussian to find the peak velocity.  
The points in the figure are approximately independent, sampling the 
kinematics at $\sim$0.2$^{\prime\prime}$.  The dashed line indicates a 
velocity of zero. The coordinates of the dynamical center (velocity = 
0 km s$^{-1}$) are approximately $\alpha$ (J2000) =13:20:35.318 and 
$\delta$ (J2000) =34:08:22.35, slightly offset from the zero offset 
position, corresponding with the map center.
}
\label{arp193_rotcurerr}
\end{figure}

We present $^{12}$CO(2-1) position-velocity diagrams for the slice 
indicated in Fig. \ref{arp193_posvel}. The corresponding rotation curve is 
shown in Fig. \ref{arp193_rotcurerr}.  The velocity at each point was obtained
by fitting a gaussian to a slice approximately two beam widths thick ($0.4\arcsec$).
We obtained the 1$\sigma$ error bars by running a 1000 trial Monte Carlo simulation whereby we
added random white noise to the map and re-fit the gaussian.  The larger error bars
farther out in the disk occur in regions with lower signal-to-noise. 
We use this rotation curve to derive 
the dynamical mass of the system and compare with the total molecular mass 
(see next section and Table \ref{arp193_mom0_clumps}).

In Figure 16 we show a comparison of our CO map 
(Fig.~\ref{arp193_mom0_clumps}) with the \HI~absorption map by Clemens and 
Alexander (2004). Contours of peak CO emission are overlaid on the 
\HI~absorption map. There are clear offsets between the peak CO emission 
and peak \HI~absorption. These spatial differences in the peak CO emission 
and peak \HI~absorption do not arise solely from errors in astrometry: no
relative shift would allow all of the peaks to line up. Comparison of our 
position-velocity maps (Fig. 14a) with the \HI\ position-velocity maps also 
shows systematic velocity differences between the CO emission and \HI\ 
absorption. In particular, the \HI~velocities do not rise quite as steeply 
as the molecular gas velocities; as discussed by Clemens and Alexander 
(2000), this is consistent with a line-of-sight distribution where most of 
the \HI\ is found at larger galactocentric distances.

\begin{figure*}
\centering
\includegraphics[totalheight=0.27\textheight]{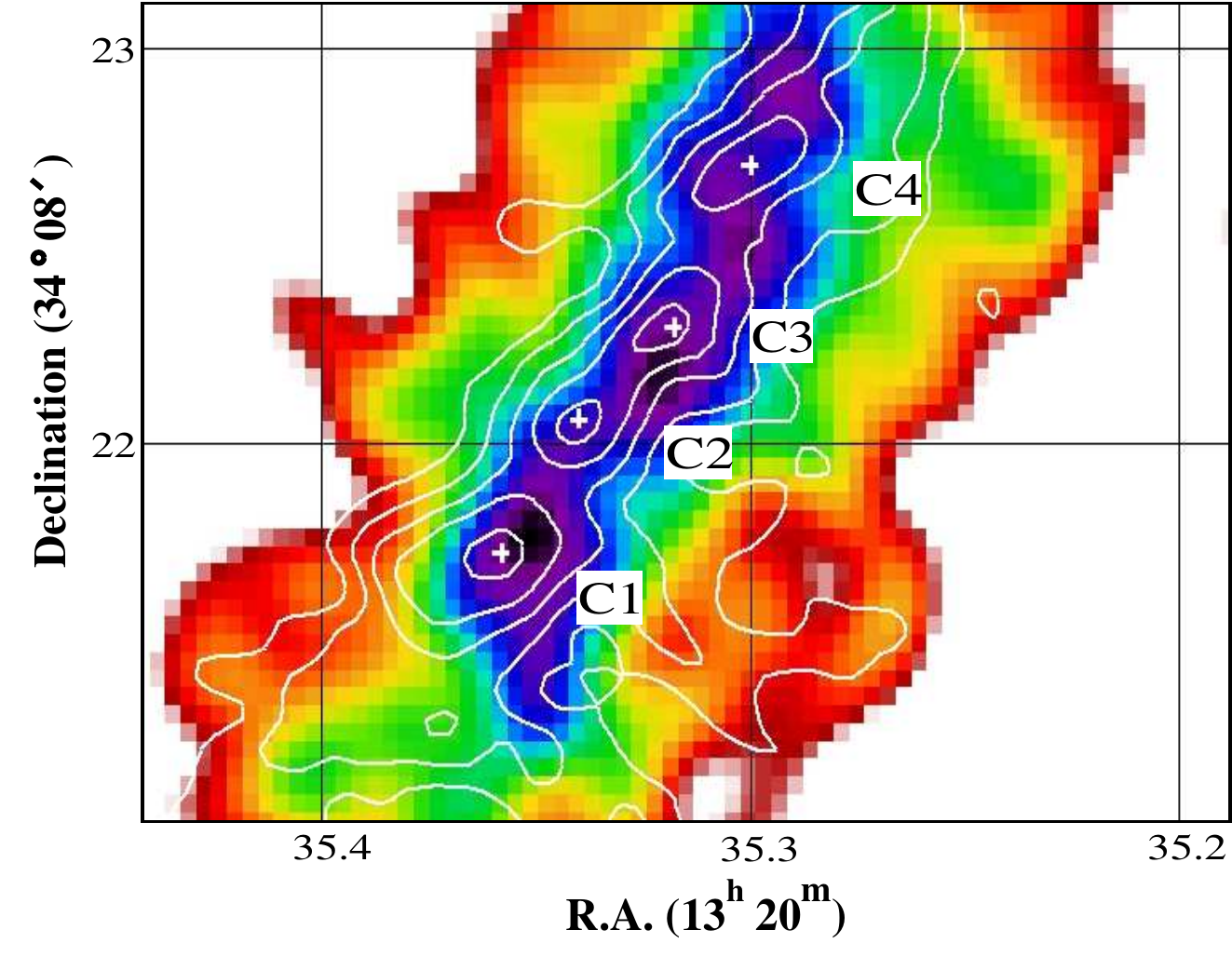}
\caption[High Resolution CO Emission and \HI~Absorption Comparison]%
{
Comparison of \HI~absorption and $^{12}$CO(2-1) emission in Arp 193.  
\HI~absorption is shown in color scale, convolved to a resolution 
of 0.$^{\prime\prime}$6, with peak of CO emission in Clumps C1$-$C4 
indicated with white cross-hairs. The overlaid contours are CO emission at 
levels of 3, 6, 9, 12 and 15 Jy km s$^{-1}$ beam$^{-1}$. The peak CO 
emission in Clumps C1 and C3 are within 0.1$-$0.2 arcseconds of the peak 
\HI\ absorption. Clumps C2 and C4 do not correspond with peaks in \HI\ 
absorption. \HI\ data is from \citet{Clemens+04}.
}\label{arp193_overlay}
\end{figure*}

\subsubsection{Molecular Gas Mass}
To compute the CO line luminosity in ${\rm K~km~s^{-1}~pc^2}$,
$L^\prime_{\rm CO}$, we use the following equation from 
\citet{1997ApJ...478..144S}:
\begin{equation}
L^{\prime}_{\rm CO} = 3.25~
\times~10^7~S_{\rm CO}~\Delta V~\nu^{-2}_{\rm obs}~D^{2}_{L}~(1+z)^{-3}.
\end{equation} 
$S_{\rm CO}\Delta{V}$ is the integrated line intensity in units of Jy
km s$^{-1}$ (see Column 5 in Table \ref{arpclump}), D$_L$ is the
luminosity distance in Mpc (98.9 Mpc for Arp 193 assuming H$_0$=71 km s$^{-1}$ Mpc$^{-1}$, 
$\Omega_m$=0.27, $\Omega_{\lambda}$=0.73, and z=0.023), and $\nu$ is the observed CO line frequency in GHz.

We compute the molecular mass using L$^{\prime}_{\rm CO}$ and the 
standard ULIRG CO-to-H$_2$ 
conversion factor, $\alpha_{CO}$, determined by DS98: 
$\alpha_{\rm CO}=0.8 {\rm M}_{\odot}({\rm K~km~s^{-1}~pc^2})^{-1}$, which
includes mass contribution from Helium by a factor of 1.36.  The 
resulting H$_2$ column and molecular surface densities are tabulated for each 
clump and for the entire region in Table 3.  The conversion factor DS98 
determined varies between 0.3 and 1.0 for other luminous and ultraluminous 
infrared galaxies, while $\alpha_{\rm CO}$ in the Milky Way is considerably
higher ($\alpha_{\rm CO}\approx4.5$ M$_{\odot}({\rm K~km~s^{-1}~pc^2})^{-1}$; 
Solomon et al. 1997 \& Bolatto et al. 2013).  
\nocite{bolatto13}
\citet{Narayan+11} and \citet{Pap+12} show that this difference in the 
conversion factor can be understood
as a result of the conditions prevalent in ULIRGs, where high gas
densities are combined with strong radiation fields and large gas
velocity gradients, as lots of molecular gas is funneled into the
central regions of merging systems.  Their findings are consistent
with the values empirically determined by DS98, mostly to avoid the
situation where the gas mass exceeds the dynamical mass of the system.

In Figure 17 we compare the dynamical mass with the molecular gas mass
of Arp~193 at a resolution of $0.2\arcsec$.  Given the limitations of
the data, the dynamical mass is approximated by inverting the rotation
curve corrected by inclination ($i=50^\circ$; DS98) assuming a
spherical mass distribution. The best fit value for the dynamical mass
based on the rotation curve (Fig.~15) is shown with the connected open
squares.  The dotted and dashed line outline the upper and lower
bounds based on propagation of error from the noise in the map, the 
fitting errors for the rotation curve and
uncertainties in the inclination of the disk.  The molecular gas mass
is indicated with the solid circles, with error bars only representing
statistical errors from the noise in the map.  
Additional sources of error in the H$_{\rm 2}$ 
mass calculation not shown in Fig. 17 include uncertainties in the 
$X_{\rm{CO}}$ factor, the distance to the source and absolute flux calibration.
Fig. 17b shows the ratio of molecular gas mass to dynamical mass,
with the dashed and dotted lines indicating the lower and upper
limits, respectively.  Out to a radius of 700 pc, the ratio approaches
a value of 0.3.  By comparison, DS98 reported a ratio of 0.19 employing the same
conversion factor out to a radius of 740 pc, a value
consistent with our lower limit on the ratio.

\begin{figure*}[htp]
\includegraphics[totalheight=0.25\textheight]{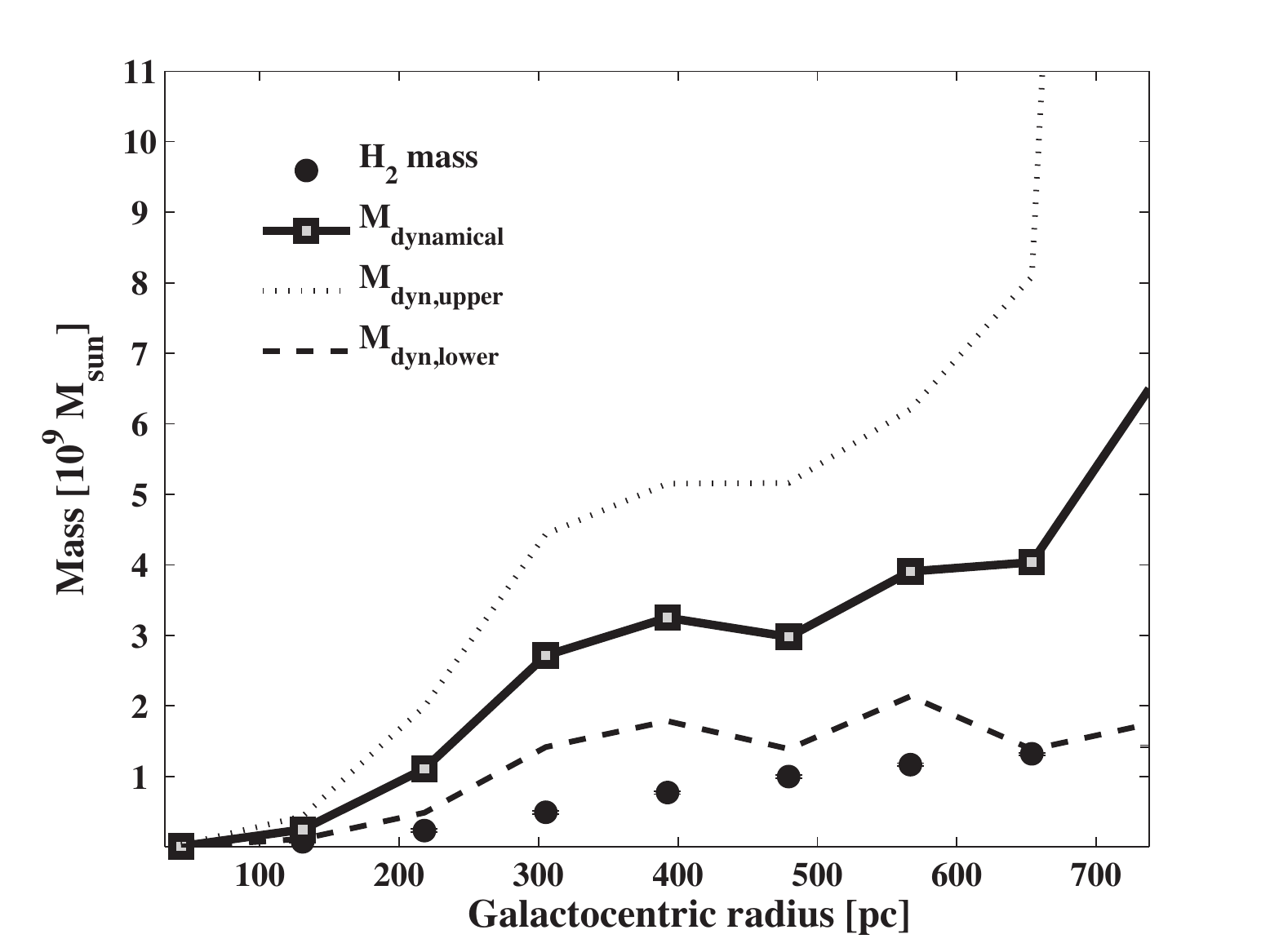}
\includegraphics[totalheight=0.25\textheight]{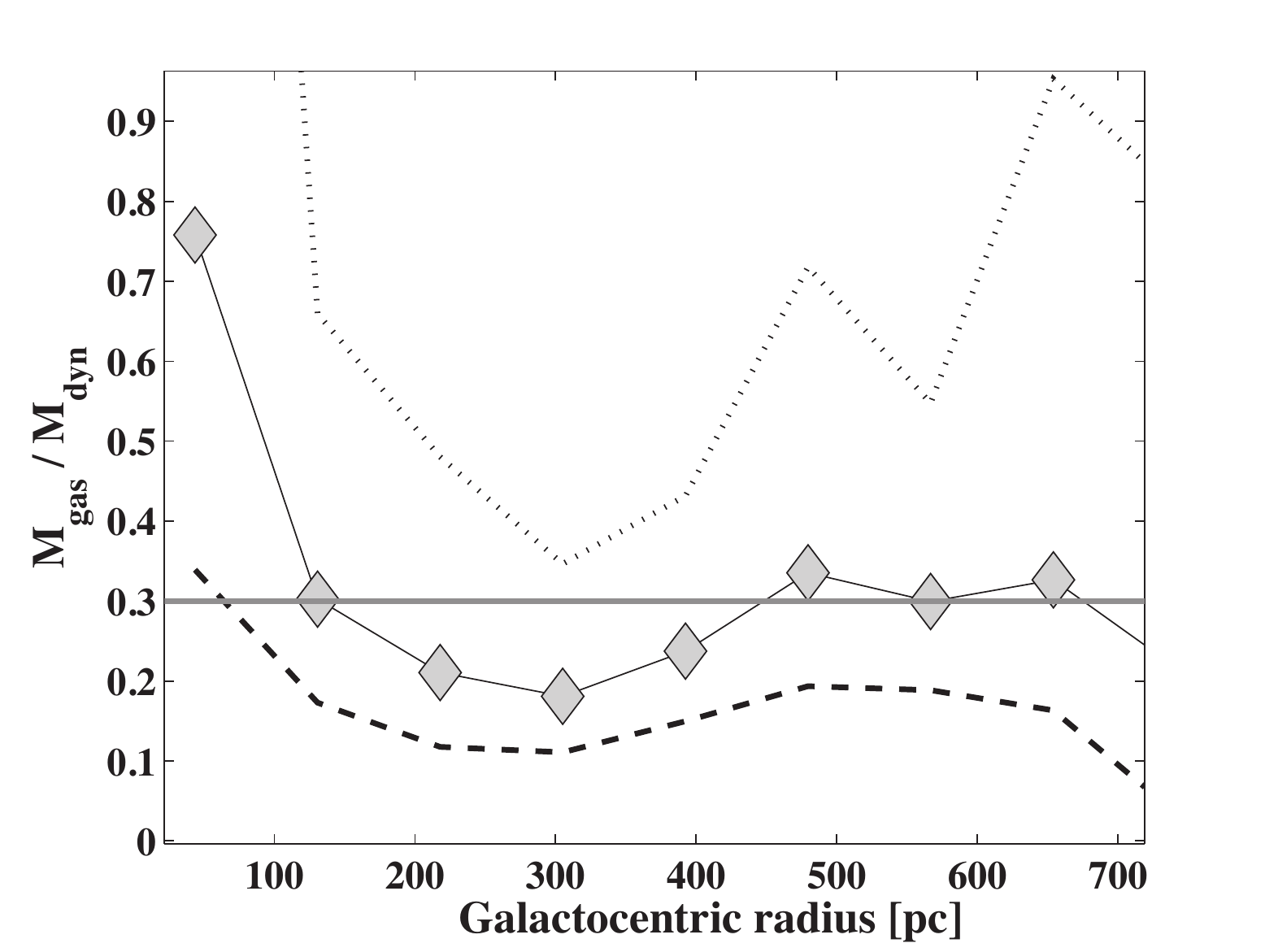}
\caption[Comparison of Molecular Gas Mass to Dynamical Mass ]%
{Comparison of dynamical and molecular masses. {\em (Left)} We compute the
dynamical mass from the derived rotation curve (see Figure
\ref{arp193_rotcurerr}), assuming an inclination  
$i=50^{\circ}$.  The gas mass is calculated from the CO line
luminosity summed over increasing radial annuli (black dots).  Our
spatial resolution is $\sim90$ pc.  
{\em(Right)} Ratio of molecular to dynamical mass.  The dotted and dashed
lines indicate the upper and lower bounds on this ratio based on
statistical errors in the mass measurements.  A ratio of 0.3 is shown
with the horizontal line.  The very high ratio in the center (inner tens of
parsecs) is artificial and due to the effect of beam smearing. The molecular gas
comprises typically about 30\% of the total mass in the disk of
Arp~193.  }
\label{mgasmdyn}
\end{figure*}

Does Arp 193 host an AGN? The column densities we observe towards 
Arp 193 (see Table 3; Column 7) are high enough to absorb even hard X-rays, resulting 
in a Compton-thick source.  Column densities of 10$^{24}$ cm$^{-2}$ (as we 
measure on scales of 80 pc) absorb X-rays with energies up to 20 keV, and 
almost all X-rays are absorbed for column densities greater than 10$^{25}$
cm$^{-2}$, likely if clumping exists within our beam.  
Teng (2010, Table 4.2) and Iwasawa et al. (2011)
summarize the X-ray properties of ULIRGs and report that Arp 193 (UGC
8387) has a point source nucleus with a hard X-ray spectrum and evidence
for far-infrared [Ne V] emission indicative of a weak AGN. More interestingly,
the soft X-ray emission is extended along the minor axis of the
molecular and stellar disk suggestive of a wind. This emission
emanates approximately from the dynamical center near Clump C3
\citep[Fig.~3;][]{Iwasawa+11}. The relative contributions of
the extreme starburst and AGN to the total observed IR luminosity in
Arp 193 remain open questions.

\begin{figure}[htp]
\centering
{
\includegraphics[trim=0cm 0cm 0cm 1cm,clip=true,angle=-90,totalheight=0.35\textheight]
                {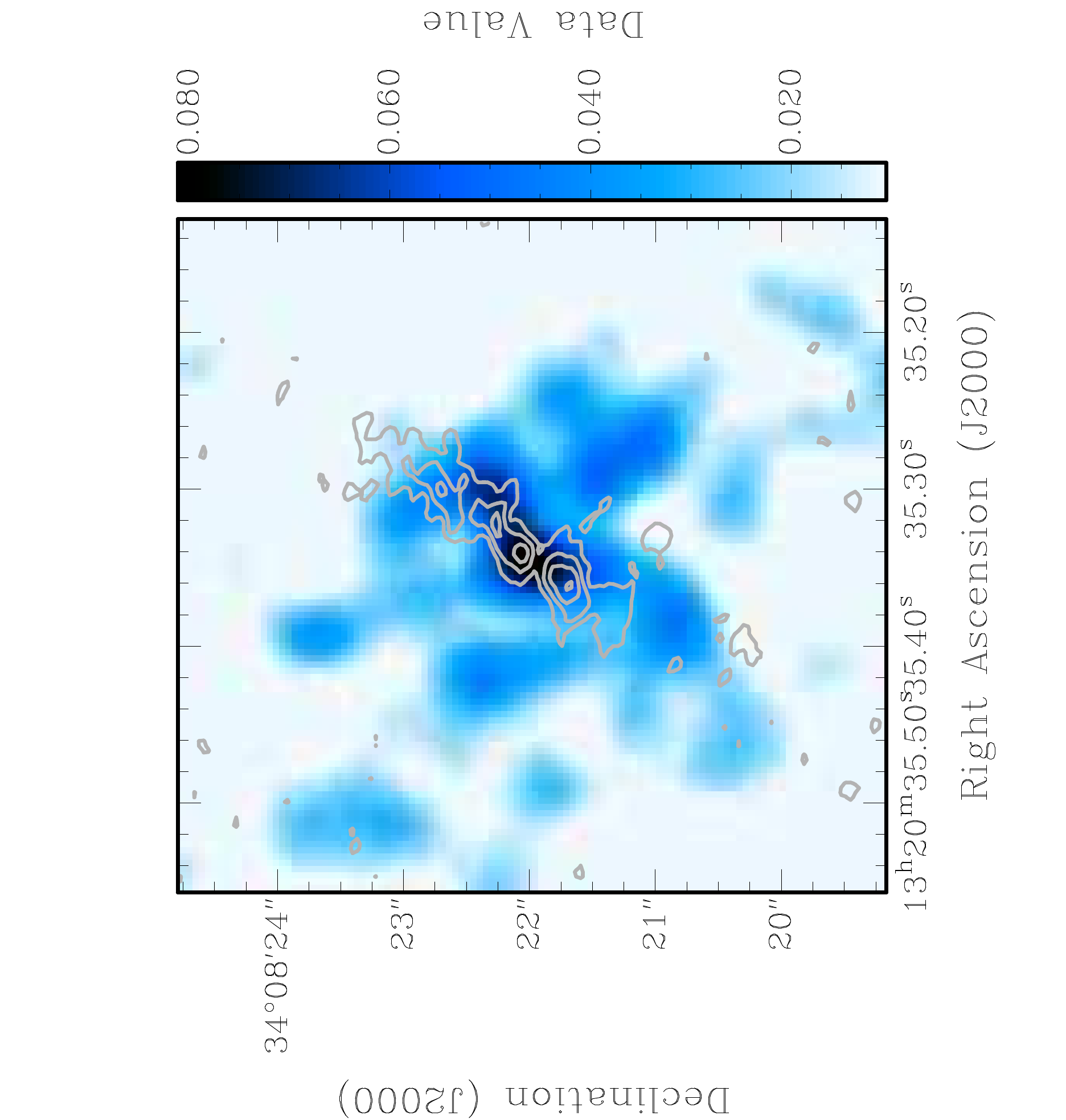}
\caption[Comparison of molecular gas with X-rays]%
{Comparison of {\it Chandra} X-ray and $^{12}$CO(2-1) emission in Arp 193.
The contours indicating the CO emission are at levels of 39, 59, 79 and 99\% peak.
The image color scale indicates X-ray emission from 0.5$-$8 keV.  The hard X-ray
emission (6$-$8 keV) is unresolved, and lies between Clumps C1 and C2.  The soft
X-ray (0.5$-$2 keV) extends orthogonal to the disk, suggestive of a galactic
wind \citep{Iwasawa+11}.  The X-ray data was smoothed with a 0.5$^\prime\prime$ Gaussian.
We note the absolute astrometric uncertainty between the 
CO and X-ray maps is 0.$^{\prime\prime}$6.}
}
\label{xray_co}
\end{figure}

\begin{deluxetable*}{cccccccc}  
\tablecolumns{8}
\tablewidth{0pt}
\tablecaption{Molecular Gas in Arp 193}
\tabletypesize{\footnotesize}
\tablehead{   
  \colhead{{\bf Clump Label}} &
  \colhead{{\bf R.A.}} &
  \colhead{{\bf Dec.}} &
  \colhead{{\bf Area}}&
  \colhead{\mbox{\boldmath{$S_{\rm CO_{2-1}}~\Delta{V}$}}}&
  \colhead{{\bf Molecular Mass}}&
  \colhead{{\bf H$_2$ Column Density}} &
  \colhead{{\bf $\Sigma_{\rm mol}$}} \\
  \colhead{} &
  \colhead{[13:20]} &
  \colhead{[34:08]} &
  \colhead{[10$^{4}$ pc$^2$]}&
  \colhead{[Jy km s$^{-1}$]}&
  \colhead{[10$^8$ M$_\odot$]}&
  \colhead{[10$^{24}$ cm$^{-2}$] }&
  \colhead{[10$^{4}$ M$_\odot$ pc$^{-2}$]} \\
  \colhead{(1)}&
  \colhead{(2)}&
  \colhead{(3)}&
  \colhead{(4)}&
  \colhead{(5)}&
  \colhead{(6)}&
  \colhead{(7)}&
  \colhead{(8)}  
}
\startdata
C1                      & 35.36 & 21.7 & 2.61   &  97.2$\pm$10.8 & 4.6$\pm$0.5   & 0.80$\pm$0.1  &  1.7$\pm$0.2 \\
C2                      & 35.34 & 22.1 & 1.34   &  49.5$\pm$6.2  & 2.3$\pm$0.3   & 0.79$\pm$0.1  &  1.7$\pm$0.2 \\
C3                      & 35.32 & 22.3 & 1.58   &  64.8$\pm$8.0  & 3.0$\pm$0.3   & 0.88$\pm$0.1  &  1.9$\pm$0.2 \\
C4                      & 35.30 & 22.7 & 1.38   &  40.5$\pm$5.3  & 1.9$\pm$0.2   & 0.63$\pm$0.1  &  1.4$\pm$0.2 \\
{\bf $\sum_{3\sigma}$}  &       &      & 23.79   &  430.1$\pm$9.1 & 20.1$\pm$0.4  & 0.39$\pm$0.01  & 0.8$\pm$0.02  
\enddata
\label{arpclump}
\tablecomments{
The flux for each clump was determined by summing the flux for pixels inside the 5$\sigma$ contour level.
To include more extended emission,  $\sum_{3\sigma}$ includes all emission at the 3$\sigma$ level.
Clump labels refer to Fig. \ref{arp193_mom0_clumps}.}
\end{deluxetable*}

We compute the ratio of \HI~ and H$_2$ column densities, using the
high resolution \HI~absorption measurements by Clemens and Alexander
(2004).  Assuming a foreground uniform screen geometry, they calculate
\HI~column densities in the range $1.7-5.5\times10^{22}$
(T$_s$/100 K) cm$^{-2}$.  With a well-mixed geometry instead, the
column density range would be larger, $4-13\times10^{22}$ (T$_s$/100
K) cm$^{-2}$.  Comparing their values to the H$_{2}$ column densities
we calculated for the regions in Table \ref{arpclump} we find
N(\HI)/N(H$_2$)$\sim 0.02$-$0.3$, with smaller values in the innermost
nuclear regions.  Clemens and Alexander report a ratio of $\sim$0.04 
(a factor of several higher assuming a well-mixed geometry).  
Although the precise result of the absorption
measurements depends on the location of the background continuum
source along the line of sight, the dominance of the molecular
phase is so large that it is extremely unlikely that it could
be due to an artifact of geometry.  

We can compare our computed surface densities (Table 3) with those in
a prototypical nuclear starburst galaxy, NGC 253. Recent mapping by
\citet{Sakamoto+11} at $\sim20$~pc resolution shows that the molecular 
emission from NGC 253 is concentrated in 5 molecular complexes, with
typical surface densities $\sim10^4$ M$_{\odot}$pc$^{-2}$ and masses
$\sim10^7$ M$_\odot$.  By comparison, the molecular complexes in Arp
193 have a similar surface densities at our 90~pc resolution, although
it is likely that clumping exists on smaller scales. Their masses,
however, are an order of magnitude larger than those of the NGC~253
complexes, $\sim10^8$ M$_\odot$ (Table \ref{arpclump}). In terms of
the total molecular mass mapped, NGC~253 is also an order of magnitude
lower ($\sim10^8$ M$_\odot$) than Arp 193 ($\sim10^9$ M$_\odot$). In
summary, each of the clumps in Figure 13 contains the molecular mass
of the entire circumnuclear starburst region in NGC~253.

\section{Conclusions}\label{pacsconclusions}
We implemented and extensively tested the paired antenna calibration
for phase correction at CARMA (C-PACS) in the extended A and B
configurations during the winter of $2009-2010$.  We used eight
paired, atmospheric calibration antennas to monitor bright quasars and
transferred phases to nearby antennas observing science targets to
correct for atmospheric phase variations on time scales of $\sim5-10$
seconds.  Analysis of the test observations of quasars and our
application to observations of Arp 193 confirm the viability of the
method.

We conclude that the angular separation between the atmospheric
calibrator and target is the single most important factor in
determining whether a C-PACS calibration is successful.  Our data show
consistent improvement in target coherence if the atmospheric calibrator is
$\lesssim6^\circ$ away from the target source.  This angular
separation limit is expected to be a function of atmospheric and site
conditions.

The C-PACS correction works well under a wide range of atmospheric
conditions.  Most interestingly, our analysis shows that C-PACS
works equally well during periods with high cloud cover and no
clouds.  Clouds have been show to dramatically hinder the performance of 
methods that rely on indirect measures of the atmospheric phase
fluctuations, such as total power or water vapor radiometry.

Ultimately, the performance we measure for the paired antenna
calibration method is limited by our implementation. In particular,
slow phase drifts between the atmospheric calibration array and the
science array are an important practical limitation for how well we
can do on faint, extended targets. The sensitivity of our atmospheric
correction antennas limits us to use calibrators that are at least
1~Jy in flux density at 30~GHz, which carries with it a limitation in sky
coverage.  Moreover, the C-PACS correction typically does not improve 
coherence for baselines shorter than 300~m, suggesting that the phase 
errors introduced
amount to at least as much as the fluctuations introduced by the
atmosphere on those scales. Finally, only eight of our science
antennas are paired with atmospheric calibration antennas. Not
surprisingly, the sampling of the atmospheric screen afforded by our
calibration correction seems to be insufficient to permit an
interpolation that provides an effective phase correction for
all the science antennas.  The lack of correction for all antennas
limits the improvement achievable in targets with extended emission,
which require to more completely sampled Fourier space.

As a science application of C-PACS, we use it to image the
very luminous infrared galaxy Arp 193 at $^{12}$CO$(2-1)$, 
improving the resolution by a
factor of $~3$ in the best published map of this galaxy. In the A
configuration of CARMA we achieved an angular resolution of
$0.18\arcsec\times0.12\arcsec$, equivalent to $84\,{\rm
pc}\times56\,{\rm pc}$ at the distance of the source.  Our
observations resolve well the rotation of the inner disk, and allow us
to measure a ratio of molecular to dynamical mass that is consistent
with 0.3 in the inner 700~pc of the object, similar to that obtained
by DS98. Comparison with the \HI\ mapping by Clemens and Alexander
(2004) shows that despite the overall resemblance there are
significant differences between the positions of the molecular peaks
and the \HI\ absorption peaks, and confirms that the gas in the inner
regions of Arp~193 is overwhelmingly in molecular form. The molecular
surface densities measured on 90~pc scales are $\sim10^4$
M$_\odot$pc$^{-2}$, similar to those reported by Sakamoto et
al. (2011) for the starburst region of NGC~253 on 20~pc scales, and
sufficient to significantly obscure a possible AGN in hard X-rays
(Teng et al. 2010; Iwasawa et al. 2011). \nocite{Iwasawa+11}
The individual clumps
(M$\sim10^8$ M$_\odot$) and the central molecular region (M$\sim10^9$
M$_\odot$), however, contain an order of magnitude more molecular gas
than the corresponding structures in NGC~253. In fact the entire
molecular mass of NGC~253 is similar to that of one of the 
molecular clumps of Arp~193 resolved in our observations.

\acknowledgements
We thank the referee for constructive comments and suggestions.
Support for CARMA construction was derived from the Gordon and 
Betty Moore Foundation, the Kenneth T. and Eileen L. Norris Foundation, 
the James S. McDonnell Foundation, the Associates of the California 
Institute of Technology, the University of Chicago, the states of 
California, Illinois, and Maryland, and the National Science Foundation. 
Ongoing CARMA development and operations are supported by the National 
Science Foundation under a cooperative agreement, and by the CARMA 
partner universities. We acknowledge support from NSF AST-0838178.
The funds for the additional hardware for the paired antennas were from a NASA
CDDF grant, an NSF-Y1 Award, and the David and Lucile Packard Foundation. 
B.A.Z. wishes to acknowledge the Department of Astronomy at the 
University of Maryland, where most of this research was conducted.
B.A.Z. also acknowledges partial support from NSF AST-1302954 (AAPF), 
NSF PHYS-1066293, and the hospitality of the Aspen Center for Physics. 
A. B. wishes to acknowledge partial support from NSF AST-0955836, a
Cottrell Scholar award from the Research Corporation for Science
Advancement, and the Humboldt Foundation. 
We thank M.~S.~Clemens and P.~Alexander for kindly providing their
reduced \HI\ data cubes for comparison and analysis.
We acknowledge the input and support in implementation of this
experiment from Owens Valley Radio Observatory staff Dave Hawkins and
Ira Snyder (correlator); Steve Scott, Andy Beard and Rick Hobbs (software
and computing); Michael Cooper, Ron Lawrence, Paul Rasmussen,
Curt Giovanine, Steve Miller and Andres Rizo (paired antenna pad 
construction and array operations); Brad Wiitala, Michael Laxen, Russ
Keeney, Stan Hudson, Mark Hodges (receivers and technical development); 
John Marzano, Gene Kahn, Mike Virgin, Mary Daniel, Lori McGraw, 
Cecil Patrick, and Terry Sepsey (general operations). 

\bibliographystyle{apj}

\end{document}